\RequirePackage[running]{lineno}
\documentclass[aps,prl,nofootinbib,twocolumn,showpacs,superscriptaddress,letterpaper,amsmath,amssymb]{revtex4-1}
\pdfoutput=1 

\usepackage{graphicx}
\usepackage{dcolumn}
\usepackage{bm}

\usepackage{atlasphysics} 
\usepackage{preprintcover}

\usepackage[colorlinks,hyperindex,breaklinks]{hyperref}
\hypersetup{linkcolor=blue,citecolor=blue,filecolor=black,urlcolor=blue} 

\def\Powheg{\textsc{Powheg}}
\def\PowhegBox{\textsc{Powheg-box}}
\def\Pythia{\textsc{Pythia}}
\def\HRes{\textsc{HRes}}
\def\Herwig{\textsc{Herwig}}

\def\ptH{\ensuremath{p^\mathrm{H}_\mathrm{T}}}
\def\yH{\ensuremath{|y^\mathrm{H}|}}

\def\njets{\ensuremath{N_{\mathrm{jets}}}}
\def\ptjet{\ensuremath{p^\mathrm{j1}_\mathrm{T}}}
\def\ttH{\ensuremath{t\bar{t}H}}
\def\bbH{\ensuremath{b\bar{b}H}}
\def\VH{\ensuremath{V\hspace{-2.0pt}H}}
\def\ifb{\mbox{fb$^{-1}$}}
\newcommand\hzz{\ensuremath{{H \rightarrow ZZ ^{*}\rightarrow 4\ell}}}
\def\hyy{\ensuremath{{H\to\gamma\gamma}}}

\newcommand\incXS{\ensuremath{33.0 \pm 5.3 \, ({\rm stat}) \pm 1.6 \, ({\rm sys})~\mathrm{pb}}}

\newcolumntype{d}[1]{D{.}{.}{#1}}

\PreprintCoverPaperTitle{ Measurements of the Total and Differential Higgs Boson Production Cross Sections Combining the \hyy\ and \hzz{} Decay Channels at $\sqrt{s}=8$~TeV\\with the ATLAS Detector}

\PreprintIdNumber{CERN-PH-EP-2015-048}

\date{\today}

\PreprintJournalName{Physical Review Letters}

\PreprintCoverAbstract{
Measurements of the total and differential cross sections of Higgs boson production are performed using 20.3~fb$^{-1}$ of $pp$ collisions produced by the Large Hadron Collider at a center-of-mass energy of $\sqrt{s} = 8$ TeV and recorded by the ATLAS detector. Cross sections are obtained from measured \hyy\ and \hzz\ event yields, which are combined accounting for detector efficiencies, fiducial acceptances and branching fractions. Differential cross sections are reported as a function of Higgs boson transverse momentum, Higgs boson rapidity, number of jets in the event, and transverse momentum of the leading jet. The total production cross section is determined to be $\sigma_{pp \to H} = \incXS $. The measurements are compared to state-of-the-art predictions.
}

\clearpage

\begin{document}

\title{
  Measurements of the Total and Differential Higgs Boson Production Cross Sections Combining the \hyy\ and \hzz{} Decay Channels at $\sqrt{s}=8$~TeV\\with the ATLAS Detector
}

\author{The ATLAS Collaboration}

\date{\today}

\begin{abstract}
Measurements of the total and differential cross sections of Higgs boson production are performed using 20.3~fb$^{-1}$ of $pp$ collisions produced by the Large Hadron Collider at a center-of-mass energy of $\sqrt{s} = 8$ TeV and recorded by the ATLAS detector. Cross sections are obtained from measured \hyy\ and \hzz\ event yields, which are combined accounting for detector efficiencies, fiducial acceptances and branching fractions. Differential cross sections are reported as a function of Higgs boson transverse momentum, Higgs boson rapidity, number of jets in the event, and transverse momentum of the leading jet. The total production cross section is determined to be $\sigma_{pp \to H} = \incXS $. The measurements are compared to state-of-the-art predictions.
\end{abstract}

\pacs{13.85.Lg,13.85.Qk,14.80.Bn}

\maketitle

This Letter presents measurements of the total and differential cross sections of inclusive Higgs boson production using $20.3~\ifb$ of $pp$ collisions produced by the Large Hadron Collider (LHC)~\cite{lhc}  at a center-of-mass energy of $\sqrt{s} = 8$ TeV and recorded by the ATLAS detector~\cite{Aad:2008zzm}. The measured cross sections probe the properties of the Higgs boson and can be directly compared to the theoretical modeling of different Higgs boson production mechanisms, such as the most recent gluon fusion (ggF) QCD calculations.
They can also be used to constrain new physics scenarios, for example using the effective field theory framework as proposed in Refs.~\cite{Giudice:2007fh,Grzadkowski:2010es,Contino:2013kra,Ellis:2014jta,Englert:2014cva}.
The analysis uses event yields measured in the \hyy\ and \hzz\ decays and detector efficiencies, both determined as described in Refs.~\cite{Diffyy,Diff4l}. The statistical uncertainties on the Higgs boson signal yields in both channels are larger than the systematic uncertainties, while the total uncertainties in the two channels are similar. Combining the analyses improves the precision of the cross-section measurements by up to 40\%, and by 25-30\% on average, with respect to the corresponding measurements in the most precise individual channel.

Distributions of the differential $pp \to H$ cross sections are reported as a function of the transverse momentum \ptH\ and the rapidity \yH\ of the Higgs boson, the jet multiplicity \njets, and the transverse momentum of the leading jet \ptjet.
The observables \ptH\ and \yH\ describe the kinematics of the Higgs boson. 
They are sensitive to perturbative QCD modeling in ggF production, which is the dominant Higgs boson production mechanism in the Standard Model (SM).
The \yH\ distribution furthermore offers a clean probe of the gluon parton distribution function (PDF) and will play a role in future PDF fits.
The \njets\ and \ptjet\ observables probe the theoretical modeling of partonic radiation in ggF production as well as the overall rate and modeling of jets in vector-boson fusion (VBF) and associated Higgs boson production (\VH\ and \ttH).
Jets produced in VBF, \VH\ and \ttH\ processes tend to have higher transverse momenta than those produced via ggF production, however the sensitivity to measuring these contributions is weak with the current amount of data.

Cross sections are extracted using a combined likelihood built from the signal yields in the \hyy\ channel and the data and background yields in the \hzz\ channel, as well as detector efficiencies, fiducial acceptances and SM branching fractions~\cite{LHCHiggsCrossSectionWorkingGroup:2011ti}. A complementary approach, using a separate likelihood,
measures the shape of the differential distributions by imposing a unity normalization constraint, which removes the implicit SM assumption on the branching fractions.
For the extraction of the signal yields and the corrections of detector efficiencies, it is assumed that the signal in both channels is due to a narrow resonance with a mass
$m_H = 125.36 \pm 0.41$~GeV as measured by the ATLAS Collaboration~\cite{CombinedMass}.
The signal yield in the \hyy\ channel is obtained from fits to the diphoton mass spectra~\cite{Diffyy},
and from the background subtracted data yield in a $m_{4\ell}$ mass window of 118 to 129~GeV for the \hzz\ channel~\cite{Diff4l}.
The fiducial acceptance in both channels~\cite{Diffyy,Diff4l} is derived using a set of Monte Carlo (MC) event generators.
\PowhegBox{}~\cite{powheg1,powheg2,powheg3}, interfaced with \Pythia8~\cite{pythia81} for showering, is used to generate ggF and VBF events, while \Pythia8 is used to simulate \VH\ and associated production with top quarks (\ttH) and $b$-quarks (\bbH).
The fiducial acceptance for events with $\yH < 1.2$ is approximately 72\% for \hyy, and 55--59\% for \hzz{}.
For higher \yH, the acceptance decreases to 35--38\% in both channels.
The fiducial acceptance is more constant as a function of the other variables and is in the range 56--62\% for the \hyy\ channel and 44--53\% for the \hzz\ channel.

After correcting the differential cross sections and normalized shapes for fiducial acceptance and branching fractions,
the corresponding measurements in both channels are found to be in good agreement with each other; \mbox{$p$-values} obtained from $\chi^2$ compatibility tests are in the range 56--99\%.

In the binned maximum-likelihood fit, the statistical uncertainty of the \hyy{} event yield is modeled using a Gaussian distribution, while the event yield in the \hzz{} channel follows a Poisson distribution due to the small sample size.
Experimental and theoretical systematic uncertainties affecting the signal yields, detector efficiencies, branching fractions and fiducial acceptance corrections are taken into account in the likelihood as constrained nuisance parameters. Nuisance parameters describing the same uncertainty sources are treated as fully correlated between bins and channels.
Systematic uncertainties on the \hyy\ and \hzz\ background estimates and efficiency correction factors, as well as the uncertainty on the integrated luminosity, are described in detail in Refs.~\cite{Diffyy,Diff4l}. The branching fraction uncertainty due to the assumed quark masses and other theoretical uncertainties are evaluated following the recommendations of Ref.~\cite{Heinemeyer:2013tqa}, considering uncertainty correlations between the \hyy\ and \hzz\ decay channels.
Uncertainties on the acceptance correction related to the choice of PDF set are evaluated by taking the envelope of the sum in quadratures of eigenvector variations of the baseline (CT10~\cite{Lai:2010vv}) and the central values of alternative (MSTW2008NLO~\cite{Martin:2009iq} and NNPDF2.3 \cite{Ball:2011mu}) PDF sets. Uncertainties on the acceptance correction associated with missing higher-order corrections are evaluated by varying the renormalization and factorization scales coherently and individually by factors of 0.5 and 2 from their nominal values, and by reweighting the  \ptH\ distribution from \PowhegBox\ to the prediction of the \HRes\ 2.2\ calculation~\cite{HRes,HRes2}. The envelope of the maximum deviation of the combined scale variations and the \ptH\ reweighting is used as the systematic variation. To account for the uncertainty in the mass measurement, the Higgs boson mass is varied by $\pm 0.4$~GeV. To assess the systematic uncertainty due to the assumption of SM cross-section fractions of the Higgs boson production modes, the VBF and \VH\ fractions are varied by factors of 0.5 and 2 from the SM prediction and the fraction of \ttH\ is varied by factors of 0 and 5. These factors are based on current experimental bounds \cite{Aad:2014eha,MassMuZZ,ATLAS:2014aga,ATLAS:2014lma,CMS:2014qaa}. The total uncertainties on the acceptance correction range from 1\% to 6\%, depending on the channel, distribution and bin . 

The total systematic uncertainties on the combined differential cross sections range from 4\% to 12\%, depending on the distribution and bin. For the kinematic variables \ptH\ and \yH, the largest systematic uncertainties on the differential cross sections are due to the luminosity and the background estimates in both channels. For the jet variables \njets\ and \ptjet, the largest systematic uncertainties on the differential cross sections are due to the jet energy scale and resolution. In the shape combination, the normalization uncertainties including luminosity, branching fractions, and efficiency uncertainties do not apply. Statistical uncertainties dominate all resulting distributions, ranging from 23\% to 75\%. 

\begin{figure}[t!]
\centering
\includegraphics[width=0.98\linewidth]{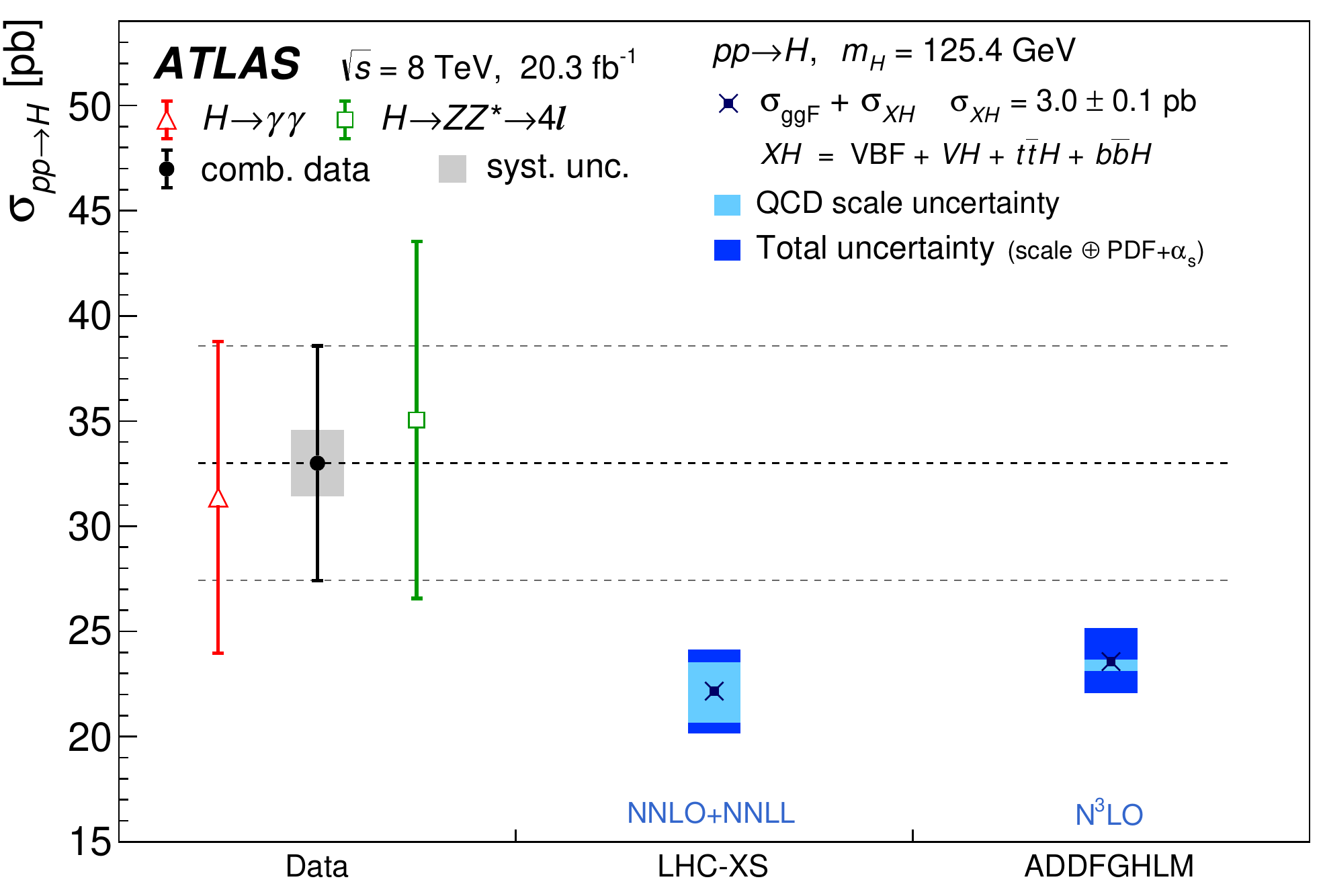}
\caption{\label{fig:inclcomp}
Measured total cross section of Higgs boson production compared to two calculations of the ggF cross section. Contributions from other relevant Higgs boson production modes (VBF, \VH, \ttH, \bbH)
are added using cross sections and uncertainties from Ref.~\cite{LHCHiggsCrossSectionWorkingGroup:2011ti}.
Details of the predictions are presented in Table~\ref{tab:Theory}. 
}
\end{figure}

\bgroup
\def\arraystretch{1.2}
\begin{table}[b!]
\centering
\caption{Summary of the ggF predictions used in the comparison with the measured cross sections.
The second column states the order in QCD perturbation theory and which threshold resummation is applied, if any.
Further details are provided in the footnotes.
All predictions are for $m_H=125.4$~GeV and $\sqrt{s}=8$~TeV.
\label{tab:Theory}}
\begin{tabular}{ll}
\hline \hline
\multicolumn{2}{c}{Total cross-section calculations} \tabularnewline
\hline
  LHC-XS~\cite{LHCHiggsCrossSectionWorkingGroup:2011ti} & NNLO+NNLL$^{\mathrm{~a,b,c}}$\\
ADDFGHLM~\cite{Anastasiou:2011pi,Anastasiou:2014vaa,Anastasiou:2014lda,Anastasiou:2015ema}  & N$^{3}$LO$^{\mathrm{~a,b,c}}$\\ \hline
\multicolumn{2}{c}{Analytical differential cross-section predictions} \tabularnewline \hline
\HRes\ 2.2~\cite{HRes,HRes2} & NNLO+NNLL$^{\mathrm{~a,e,f}}$\\
STWZ~\cite{Stewart:2013faa}, BLPTW~\cite{Boughezal:2013oha}& NNLO+NNLL$^{\mathrm{~c,d,e,g,h}}$\\
JetVHeto 2.0~\cite{Banfi:2012jm,Banfi:2012yh,Banfi:2013eda} & NNLO+NNLL$^{\mathrm{~a,c,e}}$\\ \hline
\multicolumn{2}{c}{Monte Carlo event generators} \tabularnewline
\hline
\textsc{SHERPA} 2.1.1~\cite{Gleisberg:2008ta,Hoeche:2014lxa} & $H+0,1,2$~jets @NLO$^{\mathrm{~i,j}}$\\
MG5\_aMC@NLO~\cite{mg5aMCatNLO,Frederix:2012ps} & $H+0,1,2$~jets @NLO$^{\mathrm{~i,k,l}}$\\
{\sc Powheg Nnlops}~\cite{Hamilton:2013fea,Catani:2007vq}~~ & NNLO$_{\geq 0j}$, NLO$_{\geq 1j}^{\mathrm{~e,l,m}}$\\[0.6mm]
\hline
\hline
\end{tabular}
\begin{footnotesize}
\def\arraystretch{1.0}
\begin{tabular}{ll}
$^{\mathrm{a}}$ & Considers $b$- (and $c$-) quark masses in the $gg\to H$ loop\\
$^{\mathrm{b}}$ & Includes electroweak corrections\\
$^{\mathrm{c}}$ & Based on MSTW2008nnlo~\cite{Martin:2009iq} ($\alpha_{\mathrm{s}}$ from PDF set)\\
$^{\mathrm{d}}$ & Uses $\pi^2$-resummed $gg\to H$ form factor\\
$^{\mathrm{e}}$ & NNLO refers to the total cross section\\
$^{\mathrm{f}}$ & Based on the CT10nnlo PDF set\\
$^{\mathrm{g}}$ & In the notation of Ref.~\cite{Stewart:2013faa}, this corresponds to NNLL$'$\\
$^{\mathrm{h}}$ & Includes 1-jet resummation included at NLL$'$+NLO\\
$^{\mathrm{i}}$ & Based on the CT10nlo PDF set\\
$^{\mathrm{j}}$ & Uses MEPS@NLO method and CKKW merging scheme~\cite{Hoeche:2009rj,Catani:2001cc,MEPS}\\
$^{\mathrm{k}}$ & Software version 2.2.1, NLO merged using FxFx scheme~\cite{Frederix:2012ps}\\
$^{\mathrm{l}}$ & Interfaced with \Pythia8{} for parton showering\\
$^{\mathrm{m}}$ & Uses \textsc{Minlo} method \& $y^{\mathrm{H}}$ reweighting to HNNLO~\cite{Catani:2007vq,Grazzini:2008tf,Grazzini:2013mca}.
\end{tabular}
\vspace{-2mm}
\end{footnotesize}
\end{table}
\egroup

The total \hbox{$pp\rightarrow H$} cross section is determined in the \hyy\ channel to be $31.4 \pm 7.2 \, ({\rm stat}) \pm 1.6 \, ({\rm sys})$~pb and in the \hzz\ channel to be $35.0 \pm 8.4 \, ({\rm stat}) \pm 1.8 \, ({\rm sys})$~pb.
Combining the analyses yields $\sigma_{pp \to H} = \incXS$.
Figure~\ref{fig:inclcomp} presents a comparison of these measurements with two ggF predictions to which
contributions from other relevant Higgs boson production modes (VBF, \VH, \ttH, \bbH)
are added using cross sections and uncertainties from Ref.~\cite{LHCHiggsCrossSectionWorkingGroup:2011ti}.
The LHC-XS ggF prediction, recommended in Ref.~\cite{LHCHiggsCrossSectionWorkingGroup:2011ti}, is accurate to next-to-next-to-leading order (NNLO) in QCD and utilises threshold resummation accurate to next-to-next-to-leading logarithms (NNLL).
A significant effort has been undertaken by the theory community to provide ggF cross sections beyond this precision
through various improvements in the perturbative calculations~\cite{Ahrens:2010rs,Stewart:2013faa,deFlorian:2014vta,Ball:2013bra,Bonvini:2014jma,Bonvini:2014joa}.
Recently, the ADDFGHLM group has provided a fixed-order calculation accurate to next-to-next-to-next-leading order (N$^3$LO)~\cite{Anastasiou:2011pi,Anastasiou:2014vaa,Anastasiou:2014lda,Anastasiou:2015ema}.
A PDF uncertainty of $^{+7.5}_{-6.9}$\% is assigned to the LHC-XS prediction, derived following the recommendations in Ref.~\cite{Heinemeyer:2013tqa}.
This uncertainty is increased by $^{+0.3}_{-0.1}$\% for the ADDFGHLM prediction corresponding to the change in uncertainty of the MSTW2008nnlo PDF set when changing the calculation from NNLO to N$^3$LO.
The PDF uncertainty is treated as uncorrelated with the QCD scale uncertainty.

The central value of the measured total cross section is larger than the SM predictions presented in Fig.~\ref{fig:inclcomp}.
A likelihood-ratio test statistic is used to quantify the agreement, using a bifurcated Gaussian to model the asymmetric theory uncertainties.
The resulting $p$-values are $5.5\%$ and $9.0\%$ for the agreement between data and the predictions from LHC-XS and ADDFGHLM, respectively.
The ratio of the measured cross section to the LHC-XS prediction is higher than the results presented in Refs.~\cite{Aad:2014eha,MassMuZZ,Khachatryan:2014jba},
which use an event categorization based on the expected SM yields in the different Higgs boson production modes.

\begin{figure}[t]
\centering
 \includegraphics[width=0.95\linewidth]{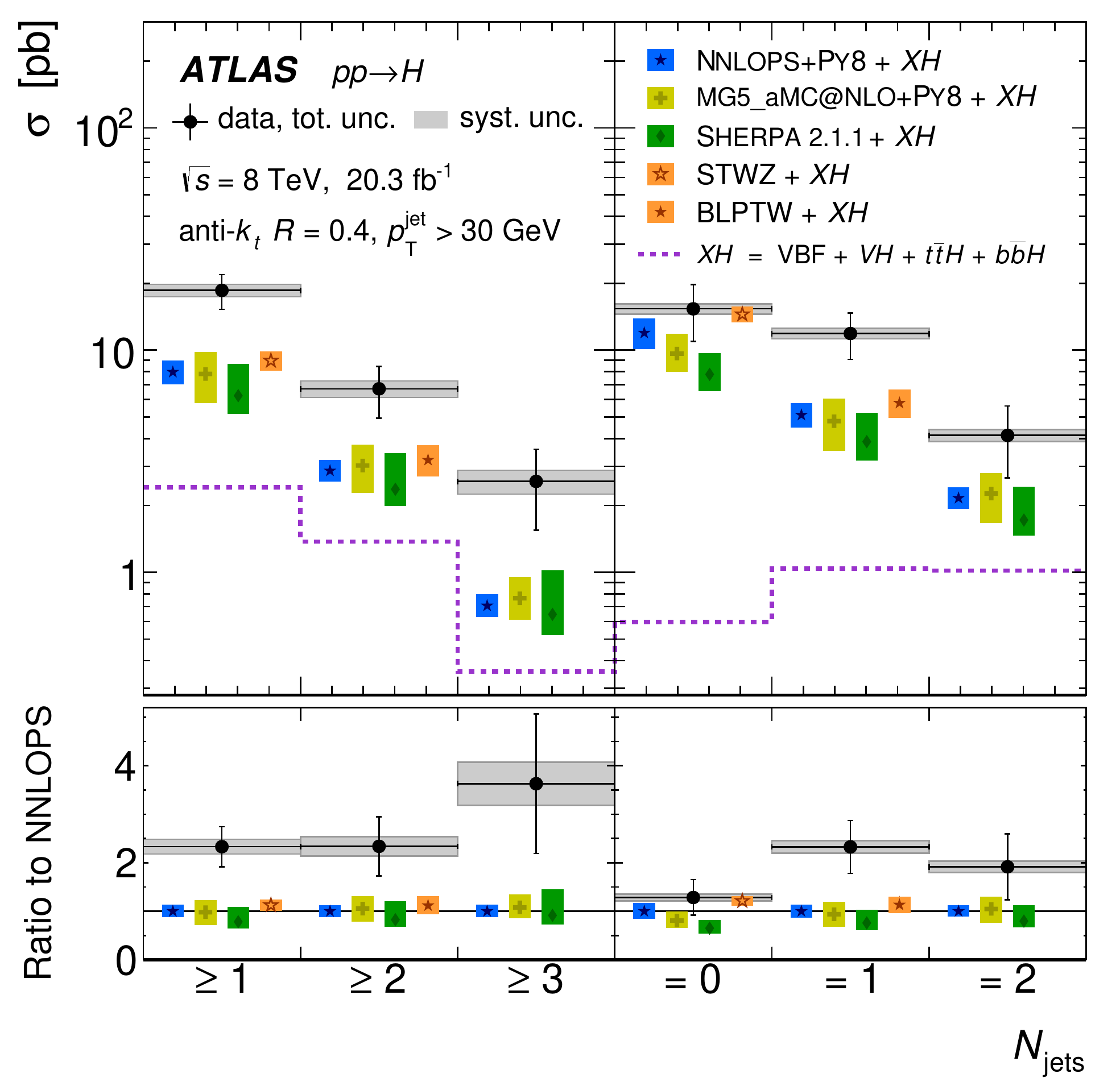}
\caption{\label{fig:jets}
Measured Higgs boson production cross sections in inclusive and exclusive jet multiplicity bins compared to different theoretical predictions (see Table~\ref{tab:Theory} for details and references). 
}
\end{figure}
\begin{figure*}[p]
\centering
 \includegraphics[width=0.44\linewidth]{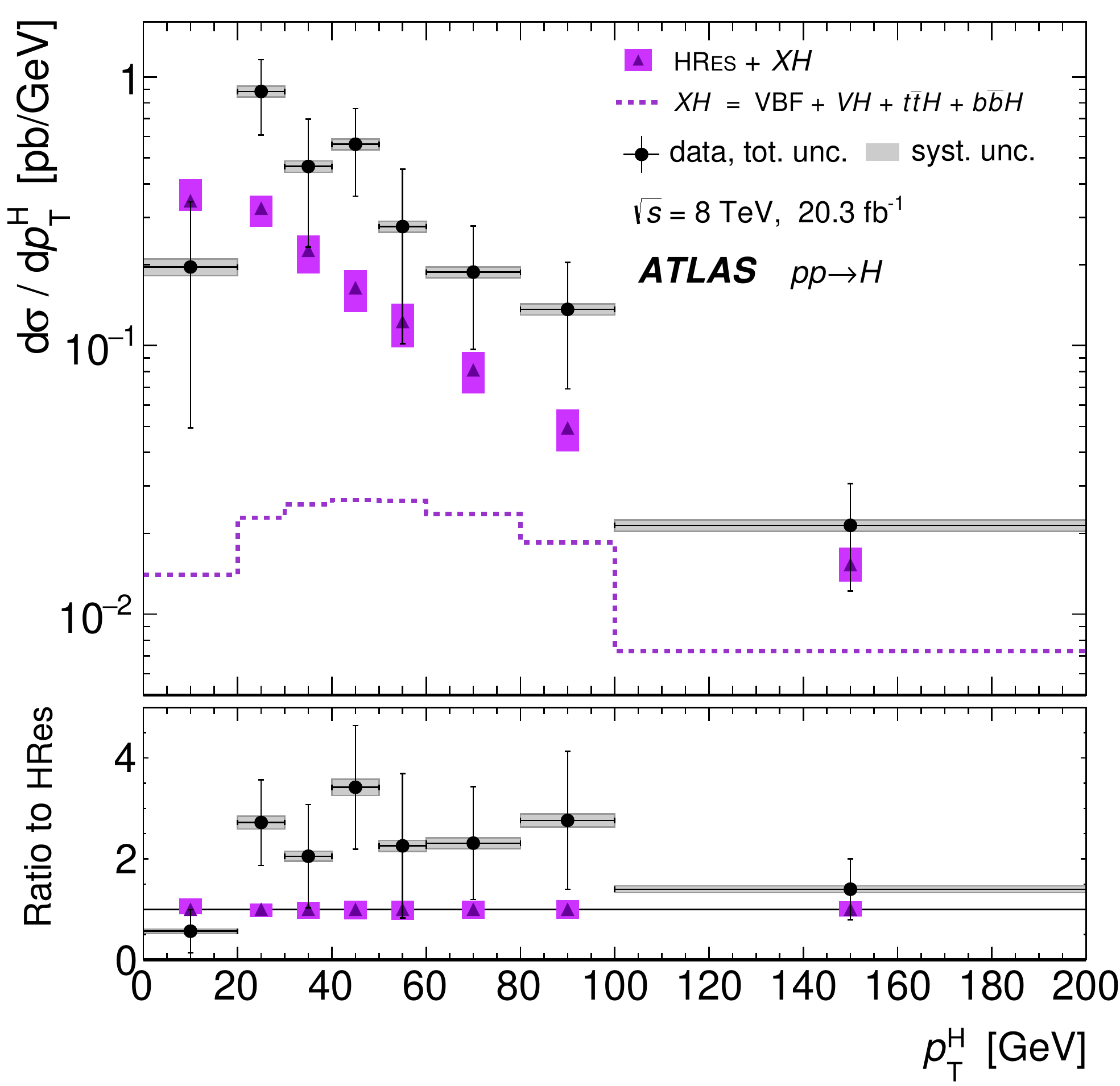}
 \includegraphics[width=0.44\linewidth]{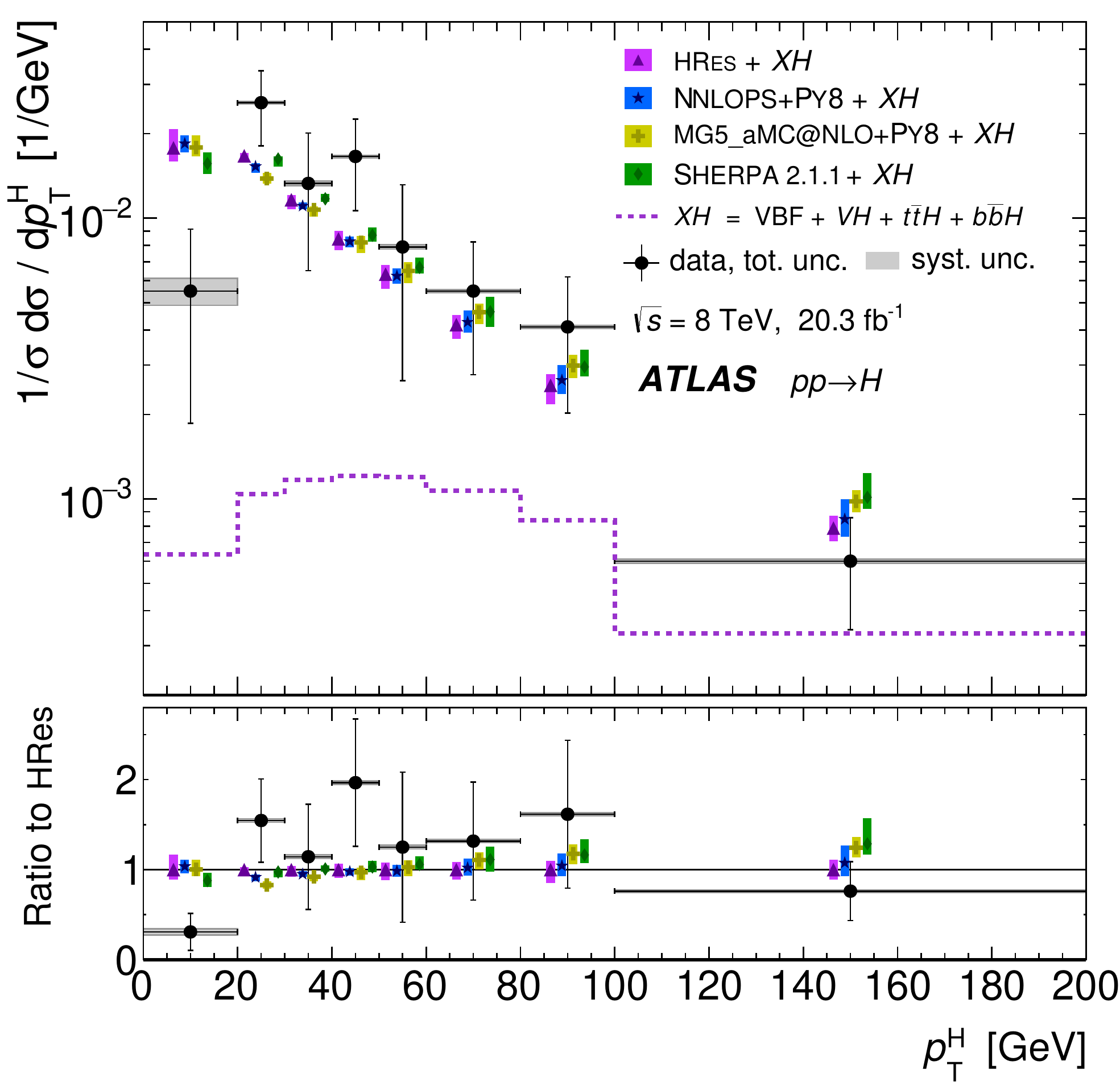}
 \includegraphics[width=0.44\linewidth]{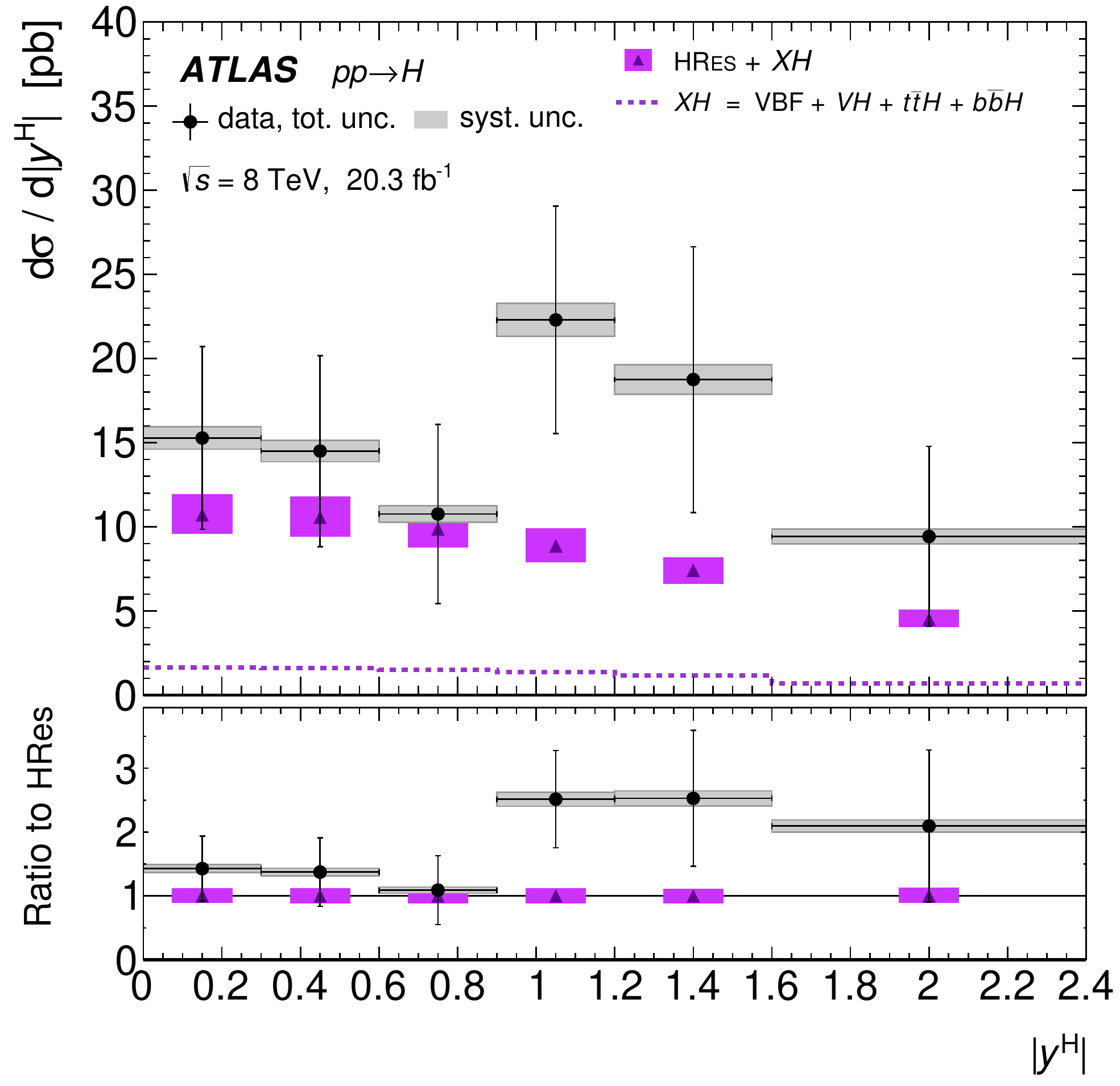}
 \includegraphics[width=0.44\linewidth]{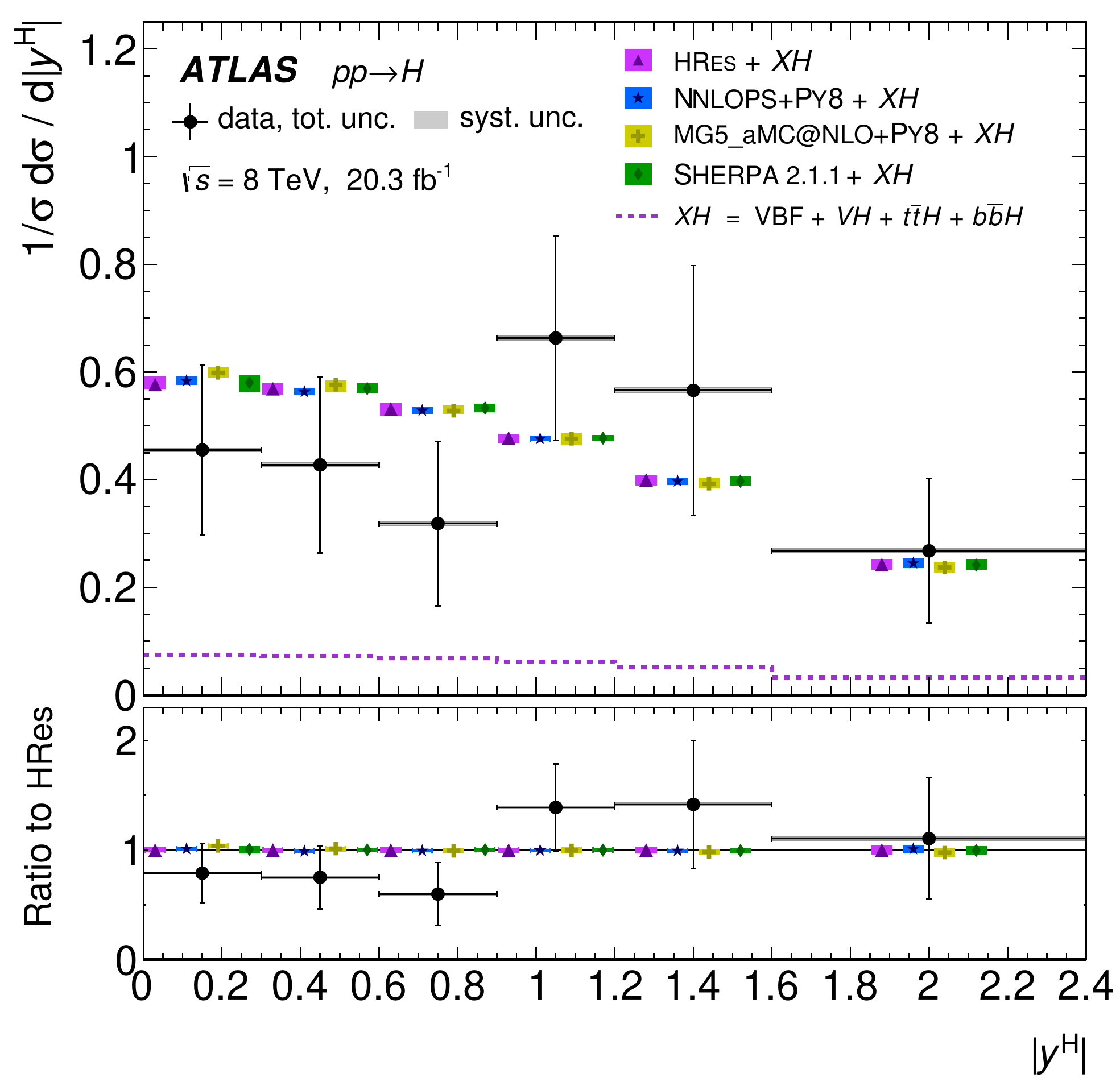}
 \includegraphics[width=0.44\linewidth]{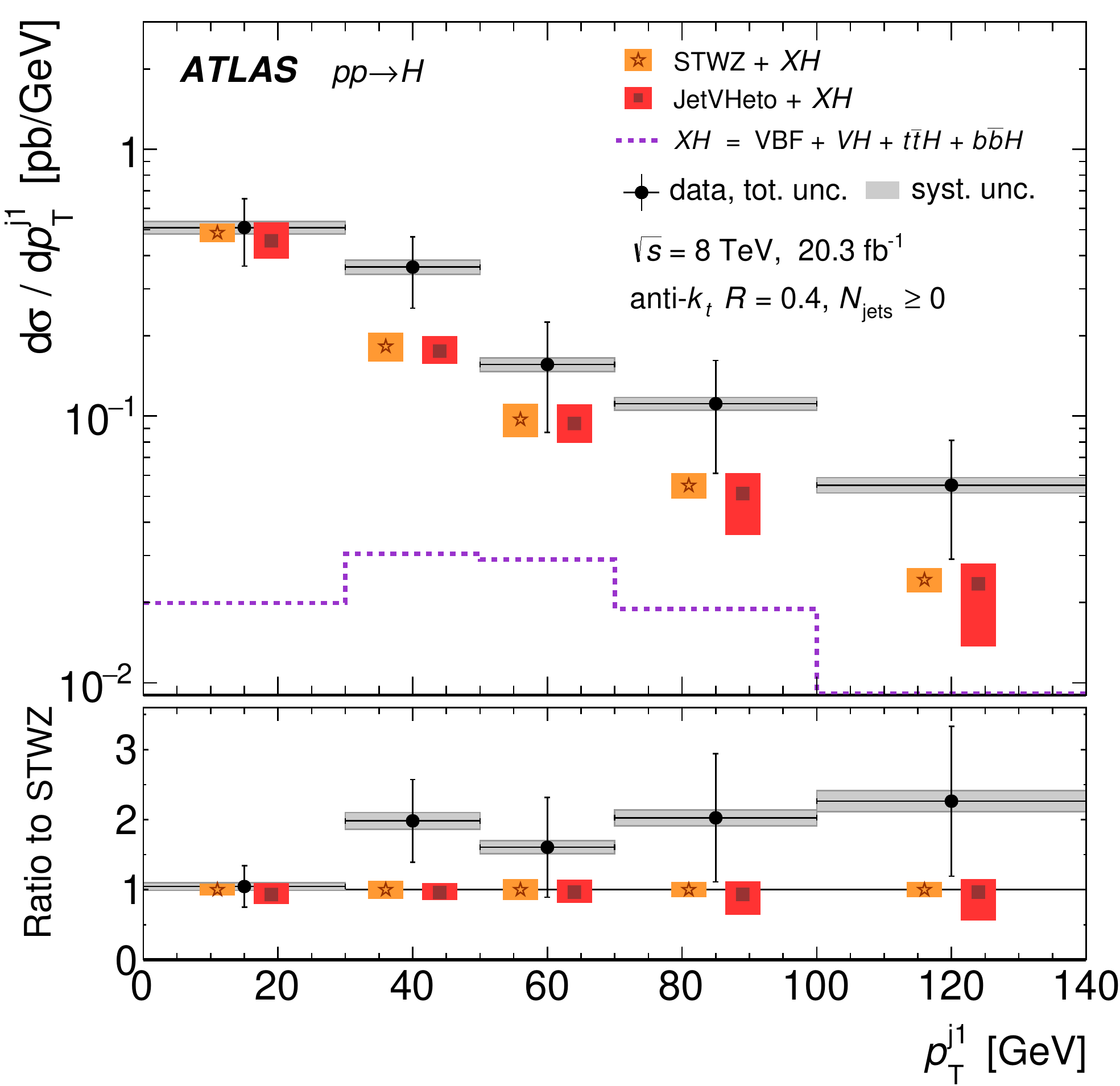}
 \includegraphics[width=0.44\linewidth]{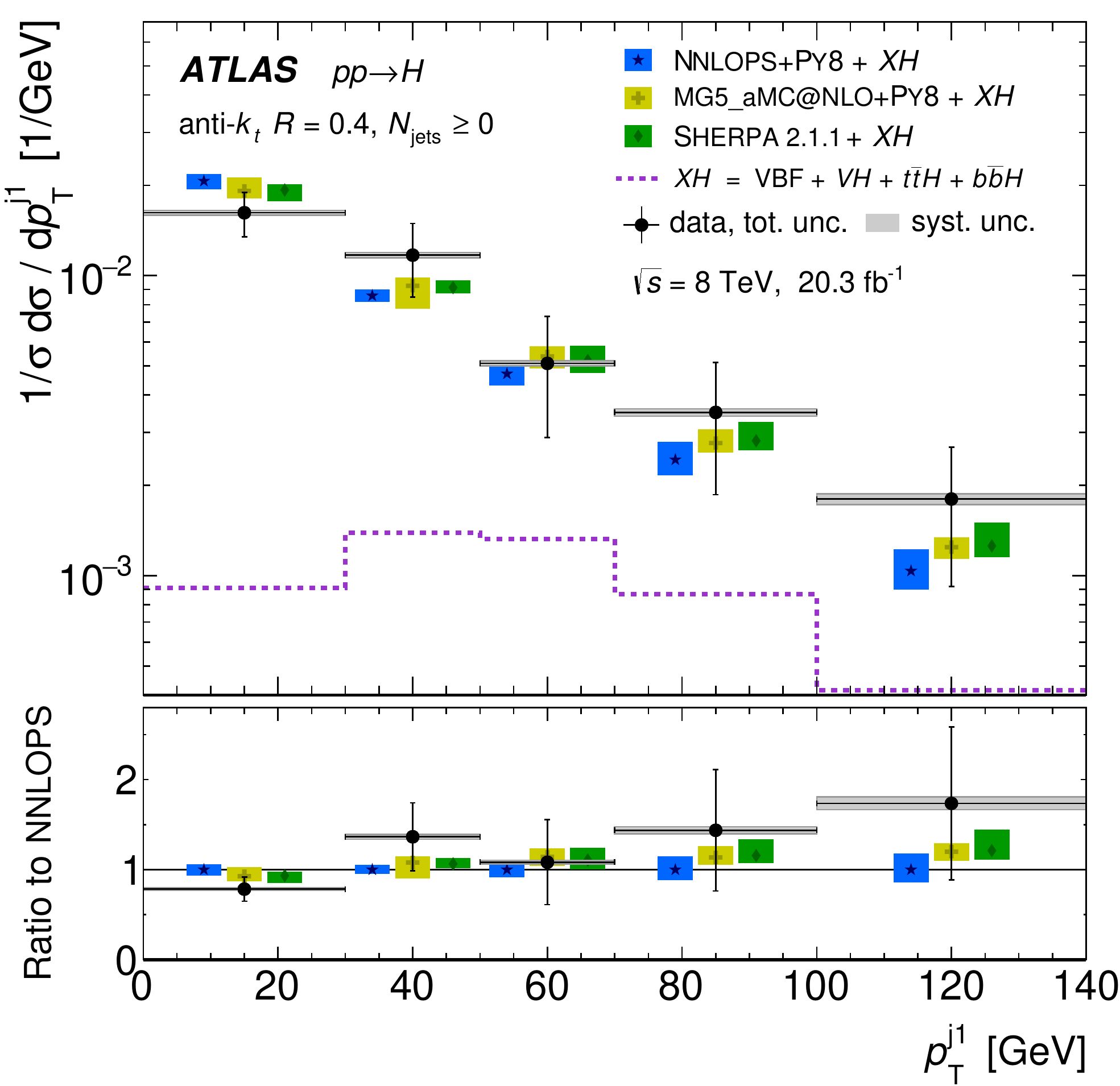}
\caption{\label{fig:XS}
Differential cross sections (left) and normalized cross-section shapes (right) for inclusive Higgs boson production measured by combining the \hyy{} and \hzz{} channels. The measured variables are the Higgs boson transverse momentum \ptH\ (top) and its rapidity \yH\ (middle), and the transverse momentum of the leading jet \ptjet{}~(bottom). The 0--30~GeV bin of the \ptjet\ distributions corresponds to events without jets above 30~GeV. Various theoretical predictions are presented, using the same bin widths as the measurement. 
}
\end{figure*}

The larger Higgs event yield observed in data motivates measurements of differential cross sections to investigate if the excess is localized to specific kinematic regions.
Figure~\ref{fig:jets} shows the comparison of the combined cross sections in different inclusive and exclusive jet multiplicity bins with state-of-the-art predictions, including NLO-accurate multi-leg (ML) merged ggF MC event generators (further details are given in Table~\ref{tab:Theory}). Jets are reconstructed using the anti-$k_t$ algorithm~\cite{AntiKt} with a radius parameter $R=0.4$~\cite{ATLASCoordinateSystem}, and are required to have $\pt>30$~GeV and $|y|<4.4$. Simulated particle-level jets are built from all particles with $c\tau > 10$~mm excluding neutrinos, electrons and muons that do not originate from hadronic decays. Photons are excluded from jet-finding if they lie inside a cone of radius $\Delta R<0.1$ of an electron or muon, and neither the photon nor lepton originate from a hadron decay.
To allow comparisons with the unfolded measurements, the analytical calculations are corrected for effects of hadronization and multiple particle interactions. These correction factors and their associated uncertainties are obtained using the \Pythia8 and \Herwig{}~\cite{herwig} MC event generators with different tunes~\cite{pythia8_AU,pythia6_PT,herwig_UE}. The total cross sections from the ML merged predictions are lower than from fully inclusive NNLO+NNLL calculations. However,
for $\njets \geq 1$, the MC predictions formally have NLO accuracy, which is the same as the analytical calculations.
Contributions from other relevant Higgs boson production modes are generated using \Powheg\ for VBF and \Pythia8 for \VH, \ttH, and \bbH, and are scaled to the cross sections in Ref.~\cite{LHCHiggsCrossSectionWorkingGroup:2011ti}.
Uncertainties are assigned to all MC predictions from QCD scale and PDF variations. The ML-merged ggF predictions also have uncertainties due to the choice of merging scale. The \textsc{SHERPA} uncertainties further include resummation scale variations.
The measured cross sections are higher than the predictions for all measured jet multiplicities. The poorest agreement between data and predictions can be found in the inclusive and exclusive \hbox{1-jet} bins, with local $p$-values ranging between 0.1\% and 3.6\%.
Normalizing the total expected cross section to the data results in an improved agreement for these bins, with local p-values ranging from 4-29\%.

The combined differential cross sections as a function of \ptH, \yH, and \ptjet\ are shown in Fig.~\ref{fig:XS} (left).
The measured \ptH{} and \yH{} distributions are compared to the \HRes\ calculation and
the \ptjet\ measurement is compared to STWZ and JetVHeto predictions.
Figure~\ref{fig:XS} (right) shows the comparisons of the normalized shapes to predictions from the MC event generators NNLOPS, SHERPA 2.1.1, and MG5\_aMC@NLO, as well as the \HRes\ calculation. The uncertainties on the predicted shapes are evaluated following the same approach  as for the differential cross-section predictions. They are derived from the impact of QCD scale, merging scale and PDF variations.
The mean of the measured \ptH{} distribution is $40.1 \pm 3.0$~GeV, while the means of the MC predictions range from 34 to 37~GeV.

The $p$-values quantifying the compatibility of the measured cross sections and  predictions range from 2\% to 26\%, and for the shapes from 8\% to 88\%. For the calculation of these values, the theory uncertainties are assumed to be Gaussian distributed and fully correlated between bins.

In conclusion, this Letter presents the first measurements of total and differential cross sections and shapes for inclusive $pp \to H$ production. The measurements were performed in the \hyy\ and \hzz\ channels using the full 2012 dataset, which consists of 20.3~fb$^{-1}$ of $pp$ collisions produced by the LHC at a center-of-mass energy of $\sqrt{s} = 8$ TeV and recorded by the ATLAS detector.
The results of the two channels are compatible and have similar precision.
The measurements indicate that the total production cross section of the Higgs boson is larger, and that it is produced with larger transverse momentum and more associated jets than predicted by the current most advanced SM calculations, however more data is needed to confirm these observations.

\vspace{5pt}
We thank CERN for the very successful operation of the LHC, as well as the
support staff from our institutions without whom ATLAS could not be
operated efficiently.

We acknowledge the support of ANPCyT, Argentina; YerPhI, Armenia; ARC,
Australia; BMWFW and FWF, Austria; ANAS, Azerbaijan; SSTC, Belarus; CNPq and FAPESP,
Brazil; NSERC, NRC and CFI, Canada; CERN; CONICYT, Chile; CAS, MOST and NSFC,
China; COLCIENCIAS, Colombia; MSMT CR, MPO CR and VSC CR, Czech Republic;
DNRF, DNSRC and Lundbeck Foundation, Denmark; EPLANET, ERC and NSRF, European Union;
IN2P3-CNRS, CEA-DSM/IRFU, France; GNSF, Georgia; BMBF, DFG, HGF, MPG and AvH
Foundation, Germany; GSRT and NSRF, Greece; RGC, Hong Kong SAR, China; ISF, MINERVA, GIF, I-CORE and Benoziyo Center, Israel; INFN, Italy; MEXT and JSPS, Japan; CNRST, Morocco; FOM and NWO, Netherlands; BRF and RCN, Norway; MNiSW and NCN, Poland; GRICES and FCT, Portugal; MNE/IFA, Romania; MES of Russia and NRC KI, Russian Federation; JINR; MSTD,
Serbia; MSSR, Slovakia; ARRS and MIZ\v{S}, Slovenia; DST/NRF, South Africa;
MINECO, Spain; SRC and Wallenberg Foundation, Sweden; SER, SNSF and Cantons of
Bern and Geneva, Switzerland; NSC, Taiwan; TAEK, Turkey; STFC, the Royal
Society and Leverhulme Trust, United Kingdom; DOE and NSF, United States of
America.

The crucial computing support from all WLCG partners is acknowledged
gratefully, in particular from CERN and the ATLAS Tier-1 facilities at
TRIUMF (Canada), NDGF (Denmark, Norway, Sweden), CC-IN2P3 (France),
KIT/GridKA (Germany), INFN-CNAF (Italy), NL-T1 (Netherlands), PIC (Spain),
ASGC (Taiwan), RAL (UK) and BNL (USA) and in the Tier-2 facilities
worldwide.

\providecommand{\href}[2]{#2}\begingroup\raggedright\endgroup

\onecolumngrid
\newpage 
\begin{flushleft}
{\Large The ATLAS Collaboration}

\bigskip

G.~Aad$^{\rm 85}$,
B.~Abbott$^{\rm 113}$,
J.~Abdallah$^{\rm 152}$,
O.~Abdinov$^{\rm 11}$,
R.~Aben$^{\rm 107}$,
M.~Abolins$^{\rm 90}$,
O.S.~AbouZeid$^{\rm 159}$,
H.~Abramowicz$^{\rm 154}$,
H.~Abreu$^{\rm 153}$,
R.~Abreu$^{\rm 30}$
% ...
\bigskip
\\
$^{1}$ Department of Physics, University of Adelaide, Adelaide, Australia\\
% ...
$^{11}$ Institute of Physics, Azerbaijan Academy of Sciences, Baku, Azerbaijan\\
% ...
$^{30}$ CERN, Geneva, Switzerland\\
% ...
$^{85}$ CPPM, Aix-Marseille Universit{\'e} and CNRS/IN2P3, Marseille, France\\
% ...
$^{90}$ Department of Physics and Astronomy, Michigan State University, East Lansing MI, United States of America\\
% ...
$^{107}$ Nikhef National Institute for Subatomic Physics and University of Amsterdam, Amsterdam, Netherlands\\
% ...
$^{113}$ Homer L. Dodge Department of Physics and Astronomy, University of Oklahoma, Norman OK, United States of America\\
% ...
$^{152}$ Institute of Physics, Academia Sinica, Taipei, Taiwan\\
$^{153}$ Department of Physics, Technion: Israel Institute of Technology, Haifa, Israel\\
$^{154}$ Raymond and Beverly Sackler School of Physics and Astronomy, Tel Aviv University, Tel Aviv, Israel\\
% ...
$^{159}$ Department of Physics, University of Toronto, Toronto ON, Canada\\
% ...
$^{s}$ Also at CERN, Geneva, Switzerland\\
% ...
\end{flushleft}


\newpage
\twocolumngrid
\section*{Supplemental Material} 
The fiducial cross section $\sigma_i$ in a given bin $i$ can be expressed as
\begin{equation}
\sigma_i = \frac{n_i}{\mathcal{L}\ \mathcal{B}\ \alpha_{i}\ c_i} = \frac{\sigma_{\mathrm{fid},i}}{\mathcal{B}\ \alpha_i},
\label{eq:xsec}
\end{equation}
where $n_i$ is the measured Higgs boson signal yield, $\mathcal{L}$ is the integrated luminosity ($20.3~\ifb$ for this analysis), $\mathcal{B}$ is the branching ratio ($0.228\%$ for \hyy{} and $0.0129\%$ for \hzz{}, $\ell=e$ or $\mu$), $\alpha_i$ is the fiducial acceptance and $c_i$ is a correction factor for detector effects, primarily accounting for reconstruction efficiency but also for bin-to-bin migration. For \hzz{}, the signal yield is defined as the number of observed events $n_{\rm data}$ in a window around the Higgs boson mass peak minus the background estimate: $n_i = n_{{\rm data},i}-n_{{\rm bkg},i}$, while for \hyy{}, the signal yield is extracted from a simultaneous signal+background fit of the $m_{\gamma\gamma}$ distribution. The correction factors for detector effects $c_i$, along with their systematic uncertainties are taken from the differential cross section measurements in the individual channels~\cite{Diffyy,Diff4l}. The differential cross section is defined as the fiducial cross section divided by the bin width.

\bgroup
\def\arraystretch{1.2}
\begin{table}[h!]
\begin{center}
\caption{Fiducial acceptance factors in percent for the \hzz\ measurement with associated uncertainties. The binning is the same as in Fig.~\ref{fig:acc_fact}.
\label{tab:zz_fid_acc}}
\footnotesize
\vspace{0.2cm}
\begin{tabular}{lccccc}
\hline\hline
Bin       &  1 &  2 &  3 &  4 & 5\\ \hline
Incl.     &  $46.7 \pm 1.1$ \\
\njets    &  $45.0 \pm 1.1$ &  $47.8 \pm 1.0$ &  $49.8 \pm 1.3$ &  $50.0 \pm 1.9$ \\
\ptjet    &  $45.0 \pm 1.1$ &  $46.2 \pm 0.8$ &  $47.7 \pm 0.8$ &  $50.3 \pm 0.7$ \\
\yH       & ~$59.2 \pm 0.9$ & ~$58.7 \pm 0.7$ & ~$57.9 \pm 0.8$ & ~$55.3 \pm 0.6$ & ~$34.8 \pm 0.8$ \\
\ptH      &  $44.4 \pm 1.0$ &  $45.2 \pm 0.9$ &  $47.6 \pm 0.7$ &  $52.9 \pm 0.7$ \\
\hline\hline
\end{tabular}
\end{center}
\end{table}
\egroup

\bgroup
\def\arraystretch{1.2}
\begin{table}[h!]
\begin{center}
\caption{Fiducial acceptance factors in percent for the \hyy\ measurement with associated uncertainties. The binning is the same as in Fig.~\ref{fig:acc_fact}.  \label{tab:yy_fid_acc}}
\vspace{0.2cm}
\begin{tabular}{lcccccccc}
\hline\hline
Bin       &  1 &  2 &  3 &  4 & 5 & 6 & 7 & 8\\ \hline
Incl.     &  $60.4 \pm 1.2$ \\
\njets    &  $62.1 \pm 1.4$ &  $59.1 \pm 0.7$ &  $58.4 \pm 1.4$ &  $55.9 \pm 3.5$ \\
\ptjet    &  $62.1 \pm 1.4$ &  $58.9 \pm 0.9$ &  $57.4 \pm 0.9$ &  $57.2 \pm 0.9$ &  $58.3 \pm 1.6$ \\
\yH       &  $71.5 \pm 1.2$ &  $71.6 \pm 1.2$ &  $71.7 \pm 1.2$ &  $71.5 \pm 1.2$ &  $67.7 \pm 1.3$ &  $37.7 \pm 0.5$ \\
\ptH      & ~$62.2 \pm 1.1$ & ~$62.2 \pm 1.2$ & ~$60.6 \pm 1.3$ & ~$59.3 \pm 1.1$ & ~$57.9 \pm 1.2$ & ~$57.5 \pm 0.9$ & ~$56.7 \pm 1.1$ & ~$58.6 \pm 1.0$ \\
\hline\hline
\end{tabular}
\end{center}
\end{table}
\egroup

\begin{figure*}[p]
\centering
\includegraphics[width=0.42\linewidth]{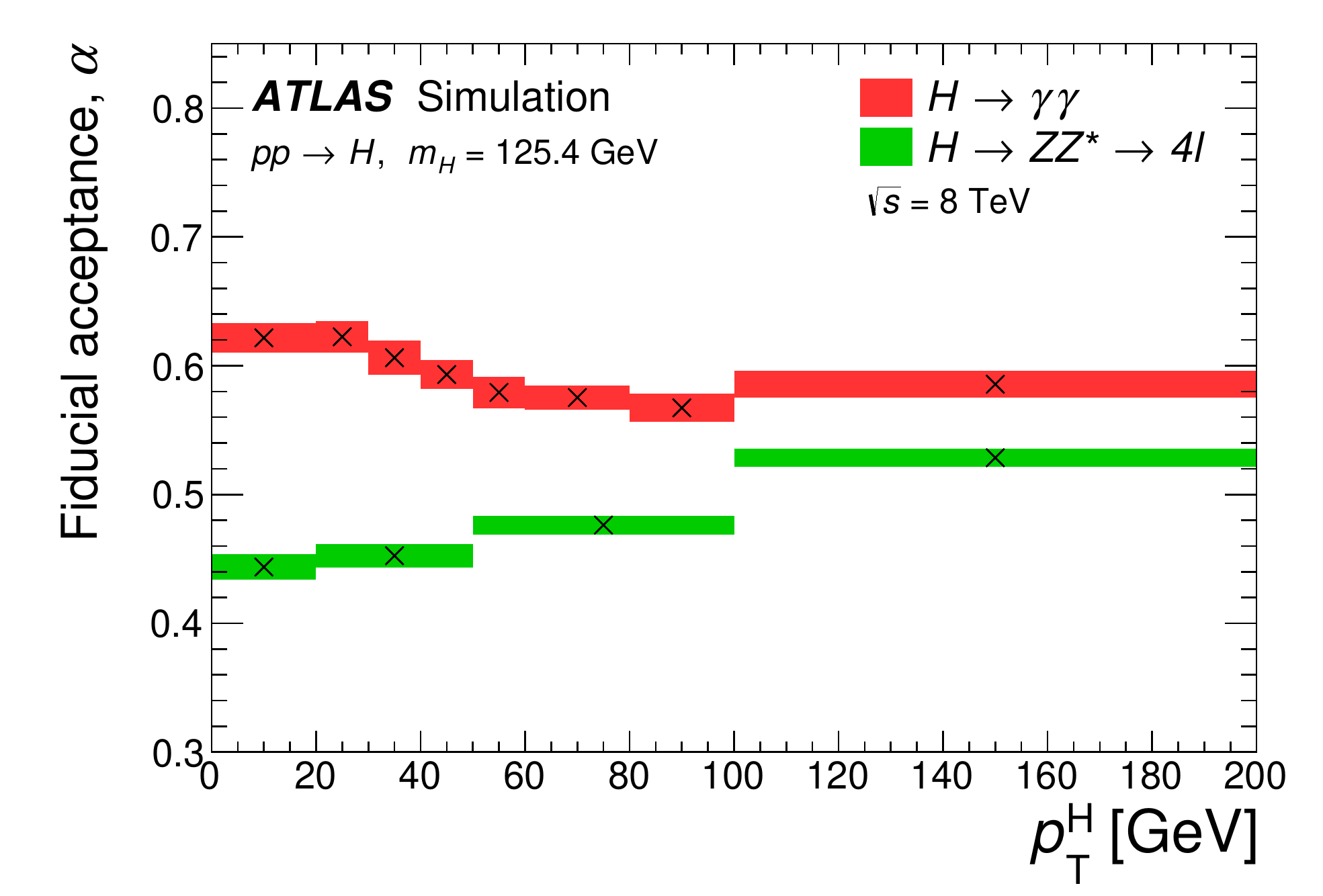}
\includegraphics[width=0.42\linewidth]{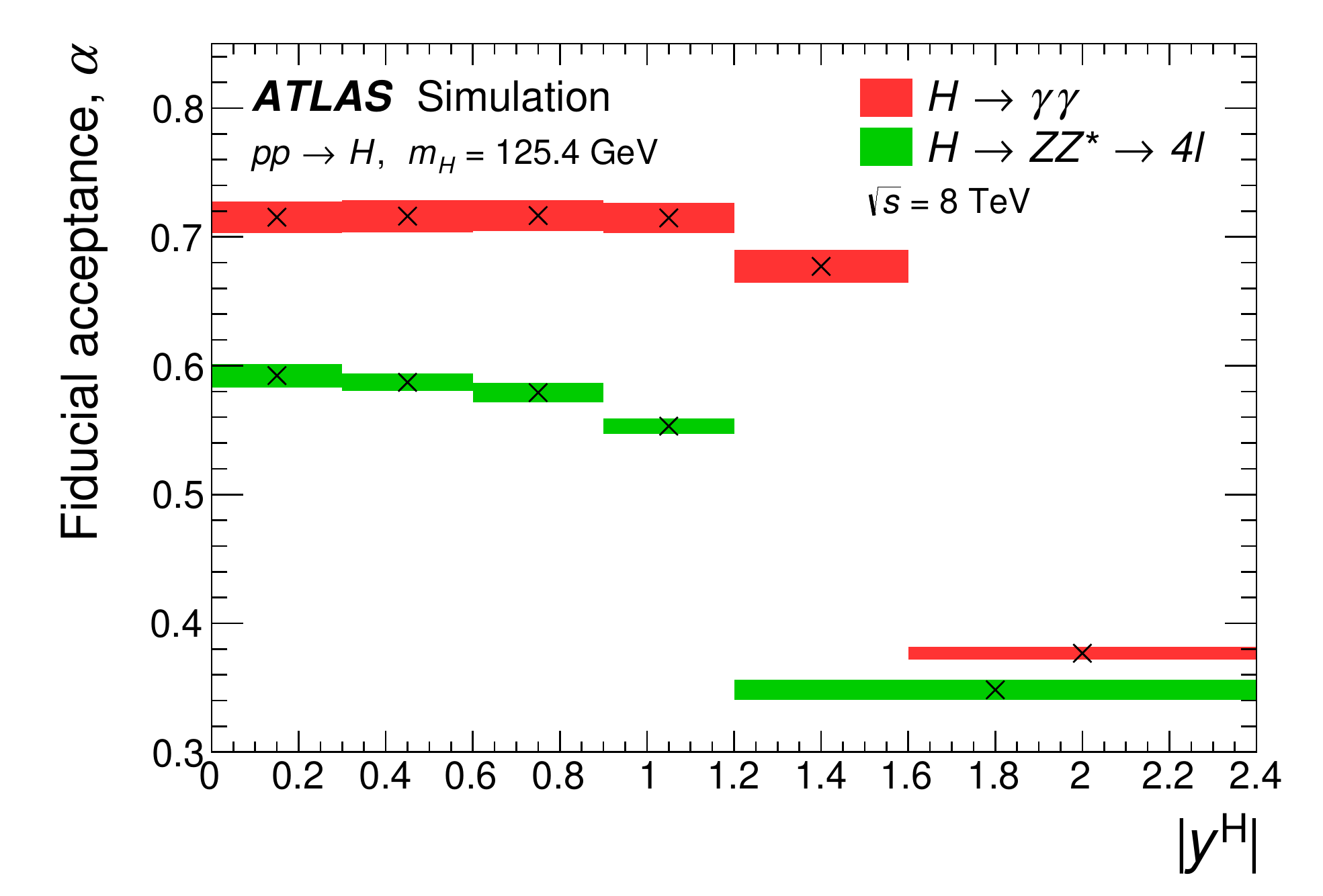}
\includegraphics[width=0.42\linewidth]{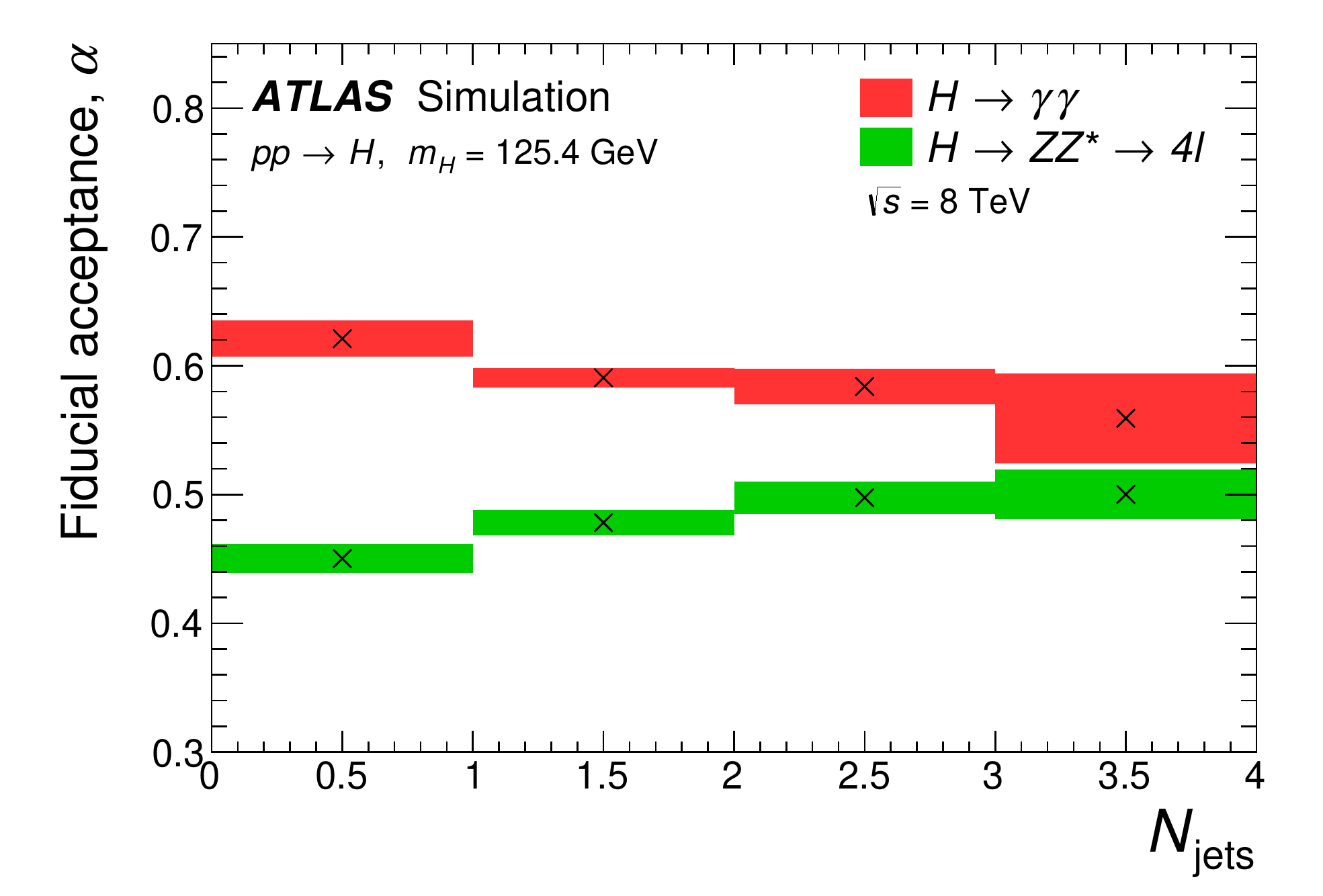}
\includegraphics[width=0.42\linewidth]{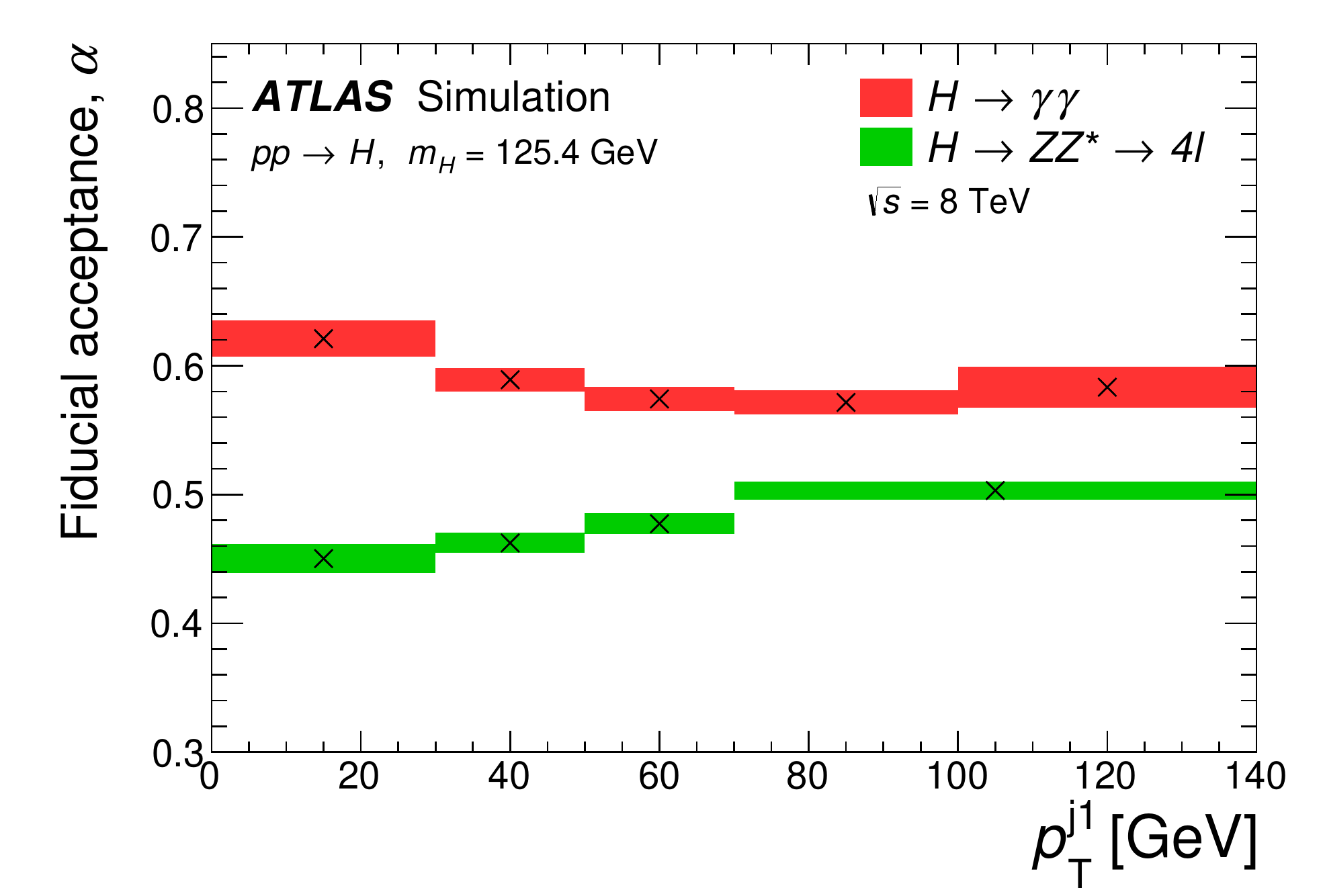}
\caption{\label{fig:acc_fact}
Fiducial acceptances mapping each measured bin of the \ptH, \yH, \njets, and \ptjet\ distributions from the inclusive phase space to the respective fiducial regions. The factors are derived using \Powheg\ for ggF and VBF production and \Pythia 8 for \VH, \ttH, and \bbH. The width of the band indicates the uncertainty from five sources: missing higher order corrections,
PDF variations, changing the Higgs boson mass and the production mode composition, and variations of the hadronization/underlying event tunes (see main text for further details).
The PDF uncertainty is the largest individual contribution to the total uncertainty.
}
\end{figure*}
\subsection{Fiducial acceptance}
For each bin $i$, the acceptance factor for each decay channel is defined as
\begin{equation}
  \alpha_i = \frac{\sigma_{{\rm fid},i}}{\sigma_{\mathrm{incl},i}}.
\label{eq:acc}
\end{equation}
The fiducial acceptances for both channels and all measured distributions are presented in Fig.~\ref{fig:acc_fact} and Tables~\ref{tab:zz_fid_acc} and~\ref{tab:yy_fid_acc}. They are based on Eq.~\ref{eq:acc} and derived using the Higgs MC samples described in the text.
For \ptH\ and \yH, $\alpha_i$ is the probability for an event to pass the fiducial requirements. The acceptance is lower for \hzz{} than for \hyy{} since it is less likely for four decay products to fulfill the fiducial requirements. For the jet variables \ptjet{} and \njets{}, an additional migration effect enters due to overlap between jets and the Higgs boson decay products, which affects the fiducial regions differently than the total phase space, where no Higgs boson decay products need to be considered.
The fiducial acceptance falls off steeply as the Higgs boson rapidity increases,  as both fiducial definitions include pseudo-rapidity requirements on the Higgs boson decay products.

\subsection{Additional figures}

Figure~\ref{fig:njets} presents the measured jet multiplicity distributions.
The lower two subfigures include the individual \hyy{} and \hzz{} measurements.
Figure~\ref{fig:xsecWchannels} presents the same six distributions as shown in Fig.~\ref{fig:XS}, but with the individual channel measurements overlaid.

\begin{figure*}[p]
\centering
\includegraphics[width=0.42\linewidth]{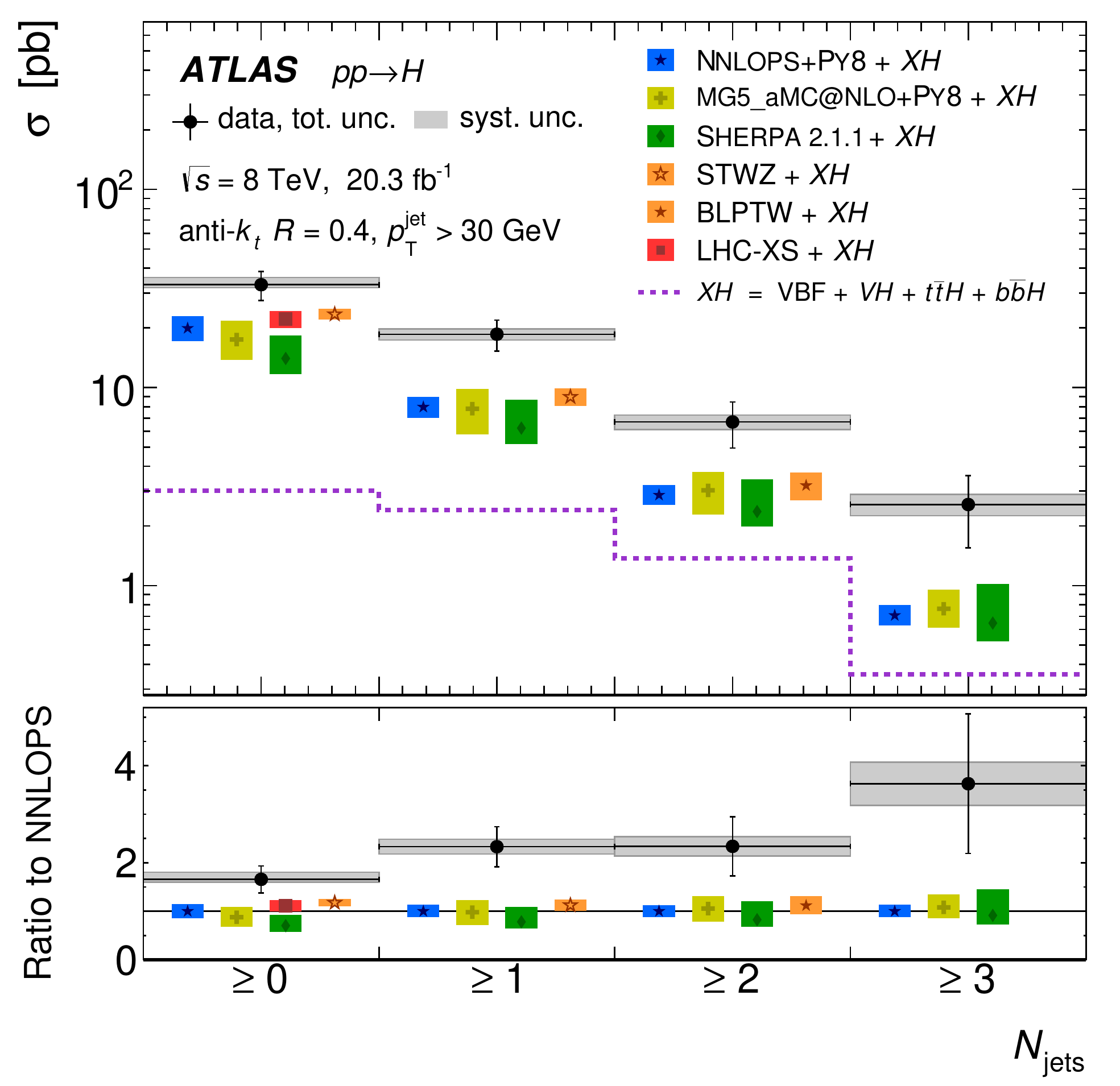}\\
\includegraphics[width=0.42\linewidth]{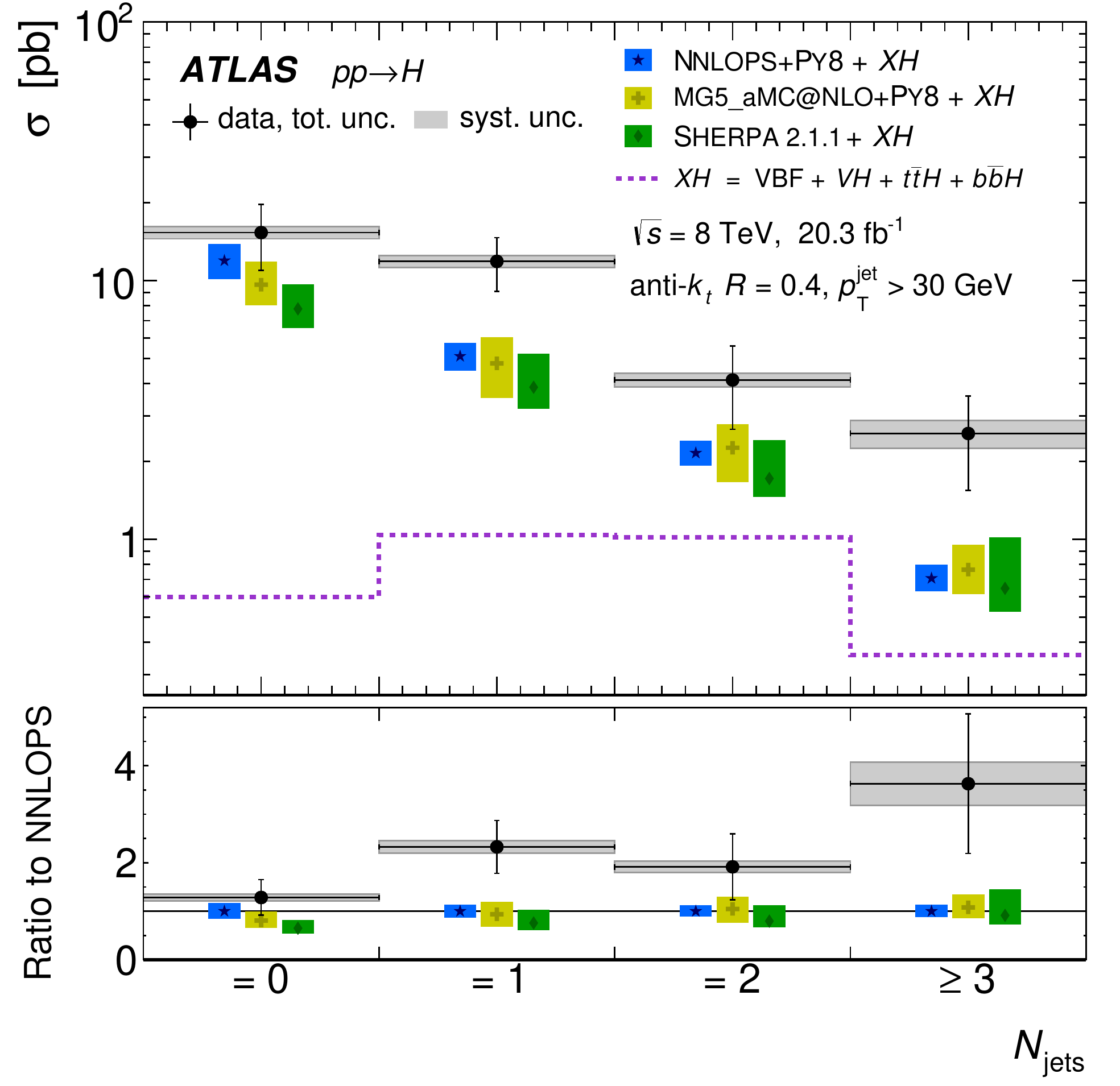}
\includegraphics[width=0.42\linewidth]{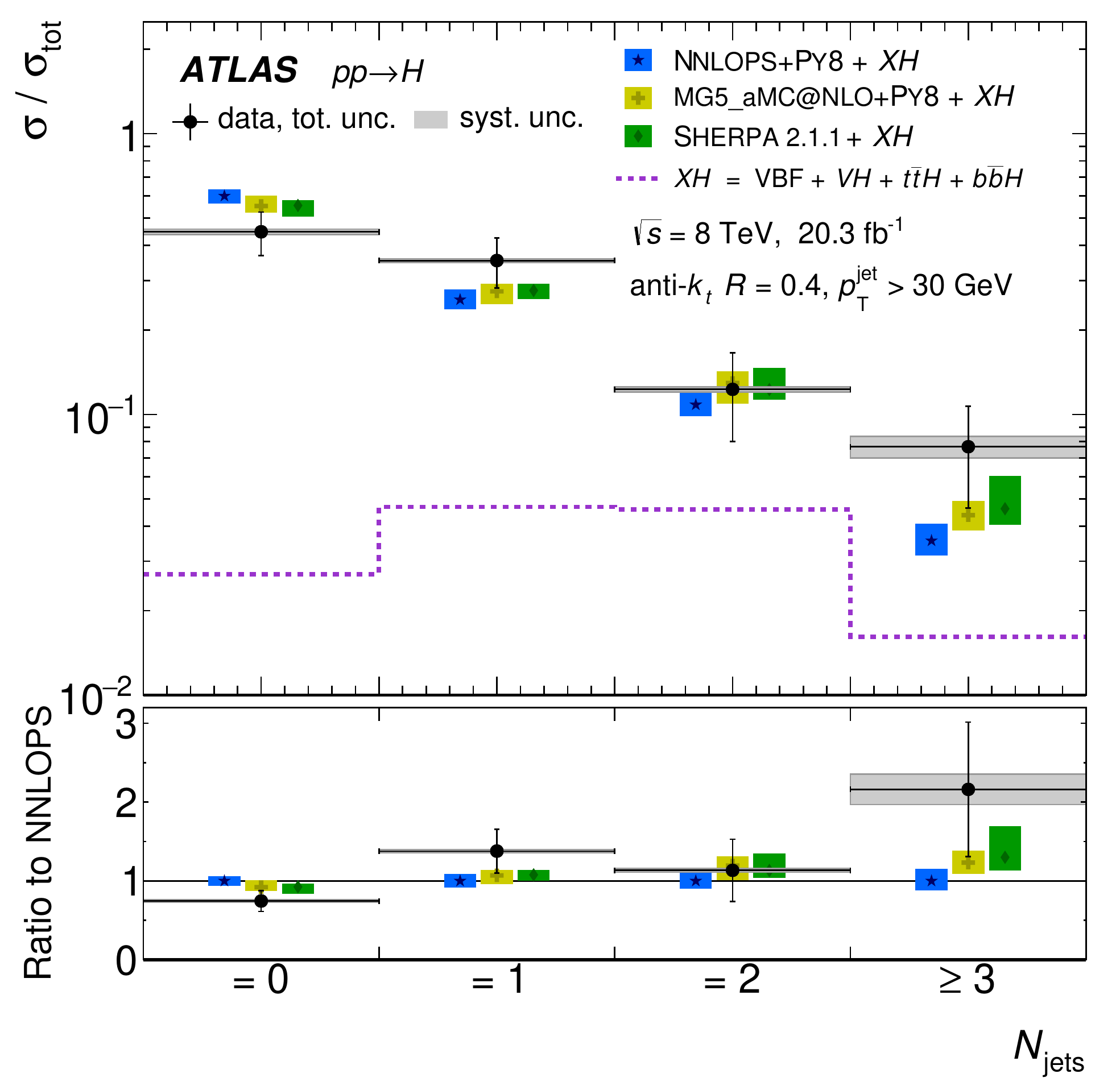}
\includegraphics[width=0.42\linewidth]{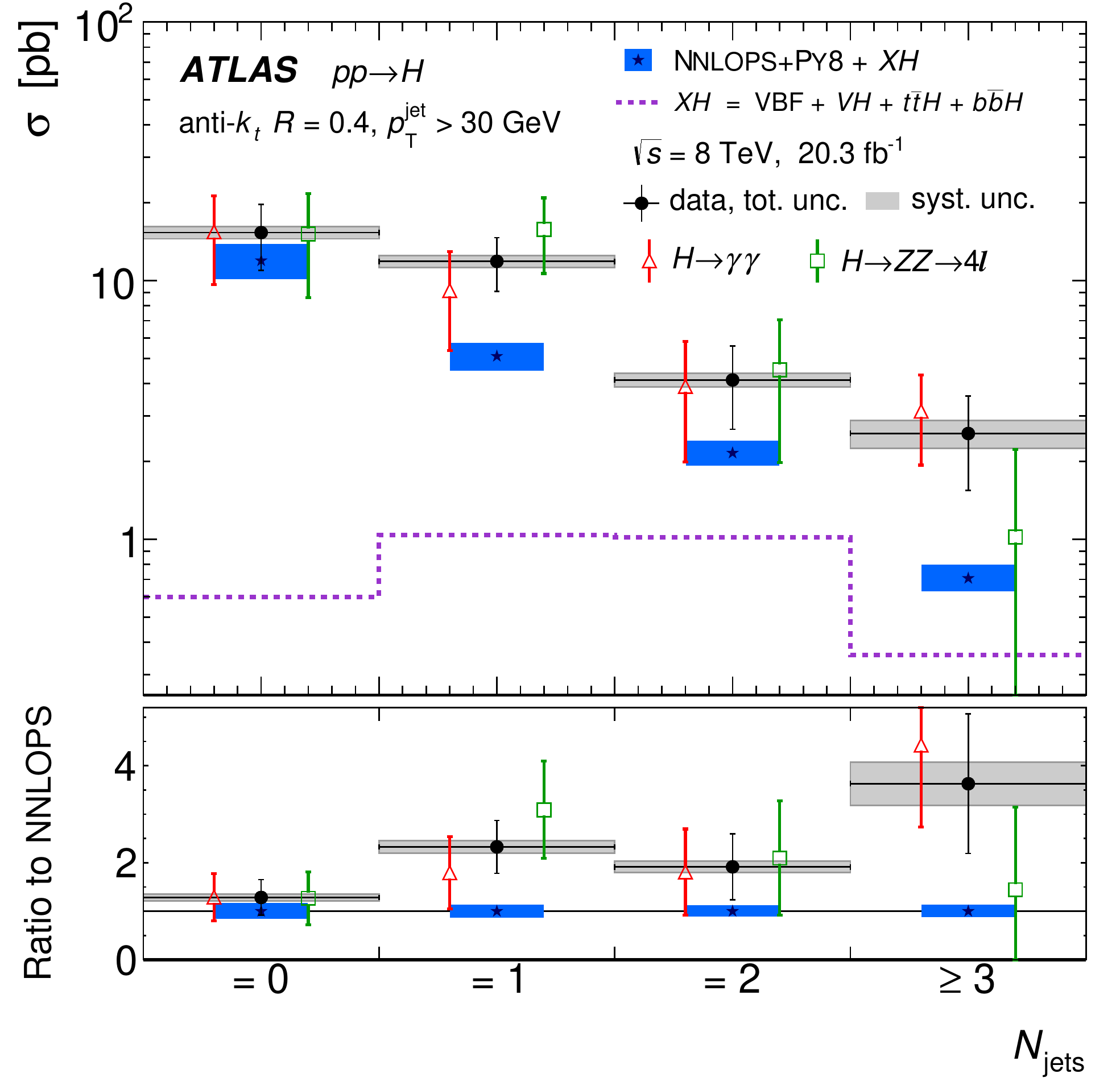}
\includegraphics[width=0.42\linewidth]{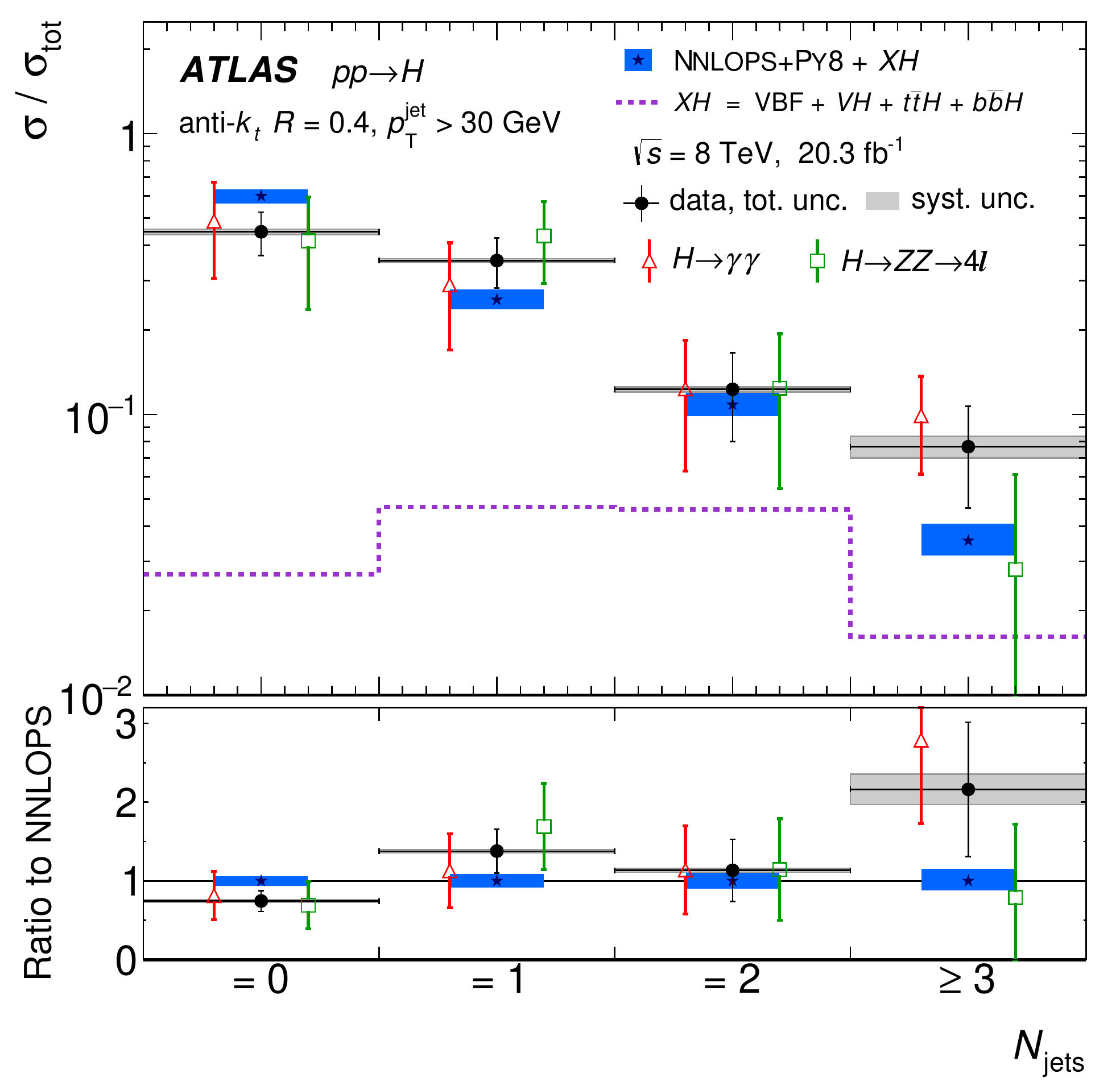}
\caption{\label{fig:njets}
Absolute and fractional cross sections in bins of jet multiplicity for inclusive Higgs boson production at $\sqrt{s}=8$~TeV measured
by combining the $\hyy{}$ and $\hzz{}$ analyses using $20.3$~\ifb{} of $pp$ collisions.
The top plot show the cross section in inclusive jet bins, while the other plots have exclusive jet binning (except for the $\geq 3$~jets bin).
In the lower two plots, the cross sections of the two channels are shown individually, defined by the fiducial cross sections corrected for acceptance and branching ratio (see Eq.~\ref{eq:xsec}).
These cross sections have partially correlated systematic uncertainties that are considered in the combined measurement.
}
\end{figure*}

\begin{figure*}[p]
\centering
\includegraphics[width=0.42\linewidth]{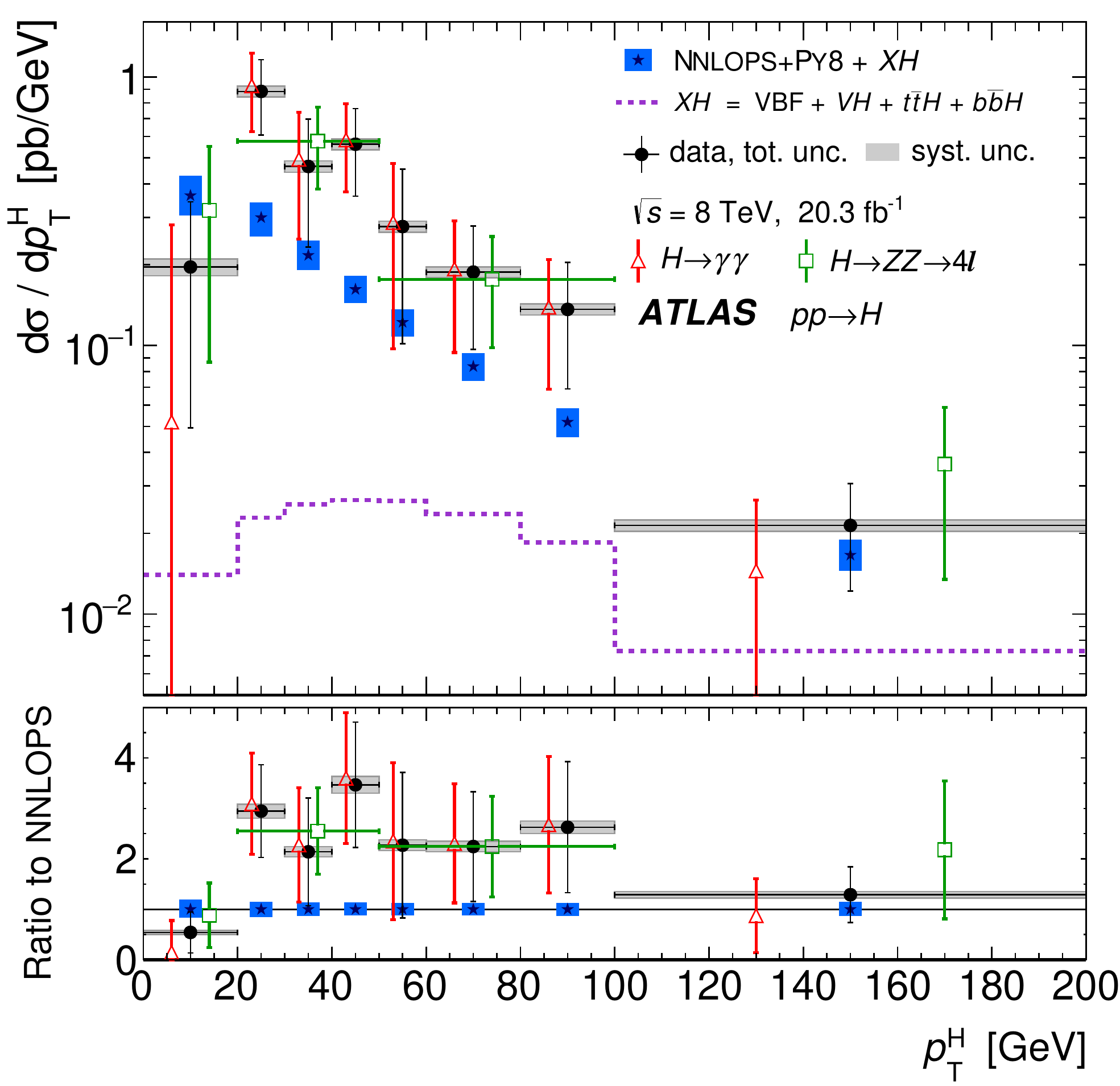}
\includegraphics[width=0.42\linewidth]{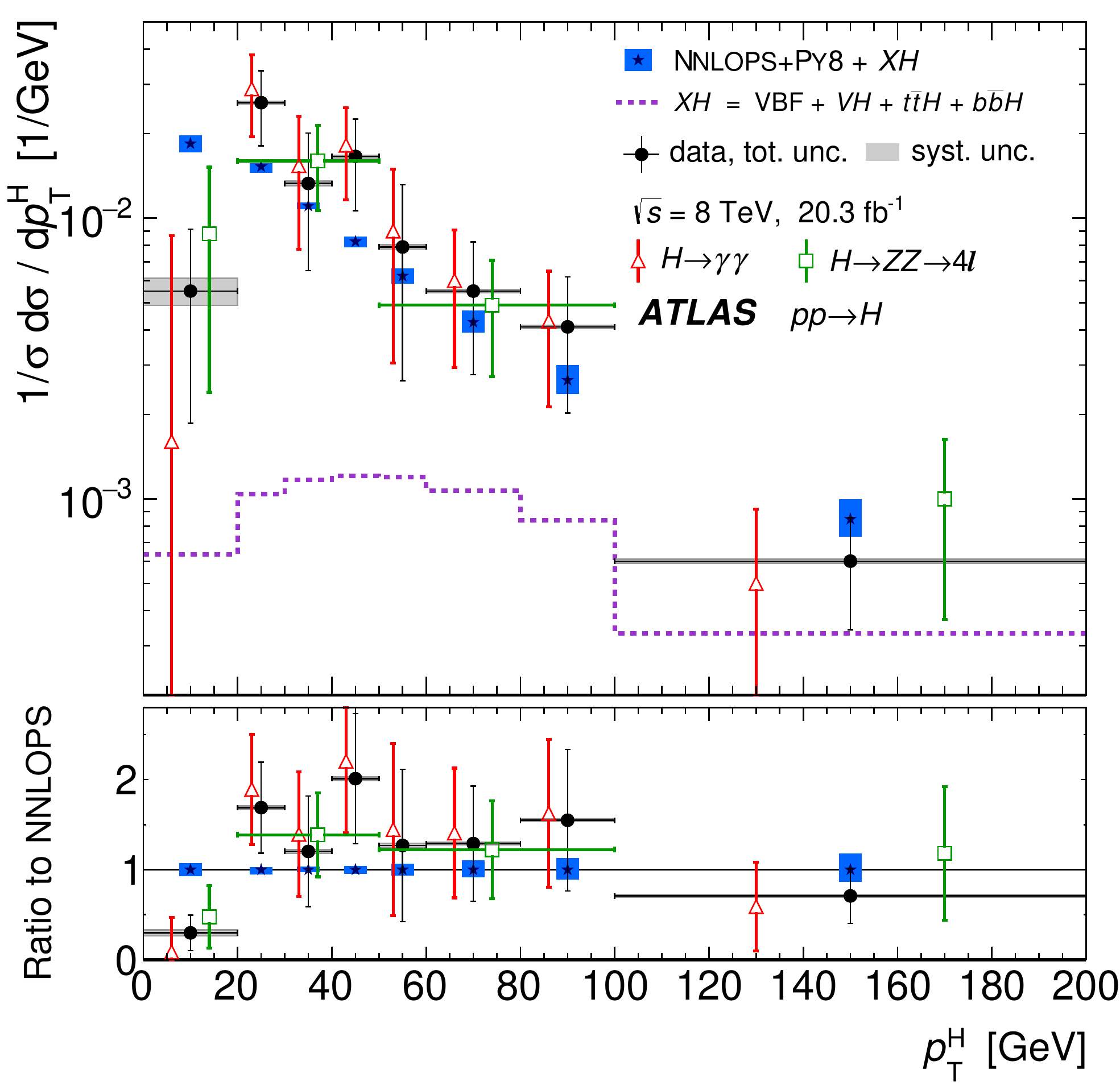}
\includegraphics[width=0.42\linewidth]{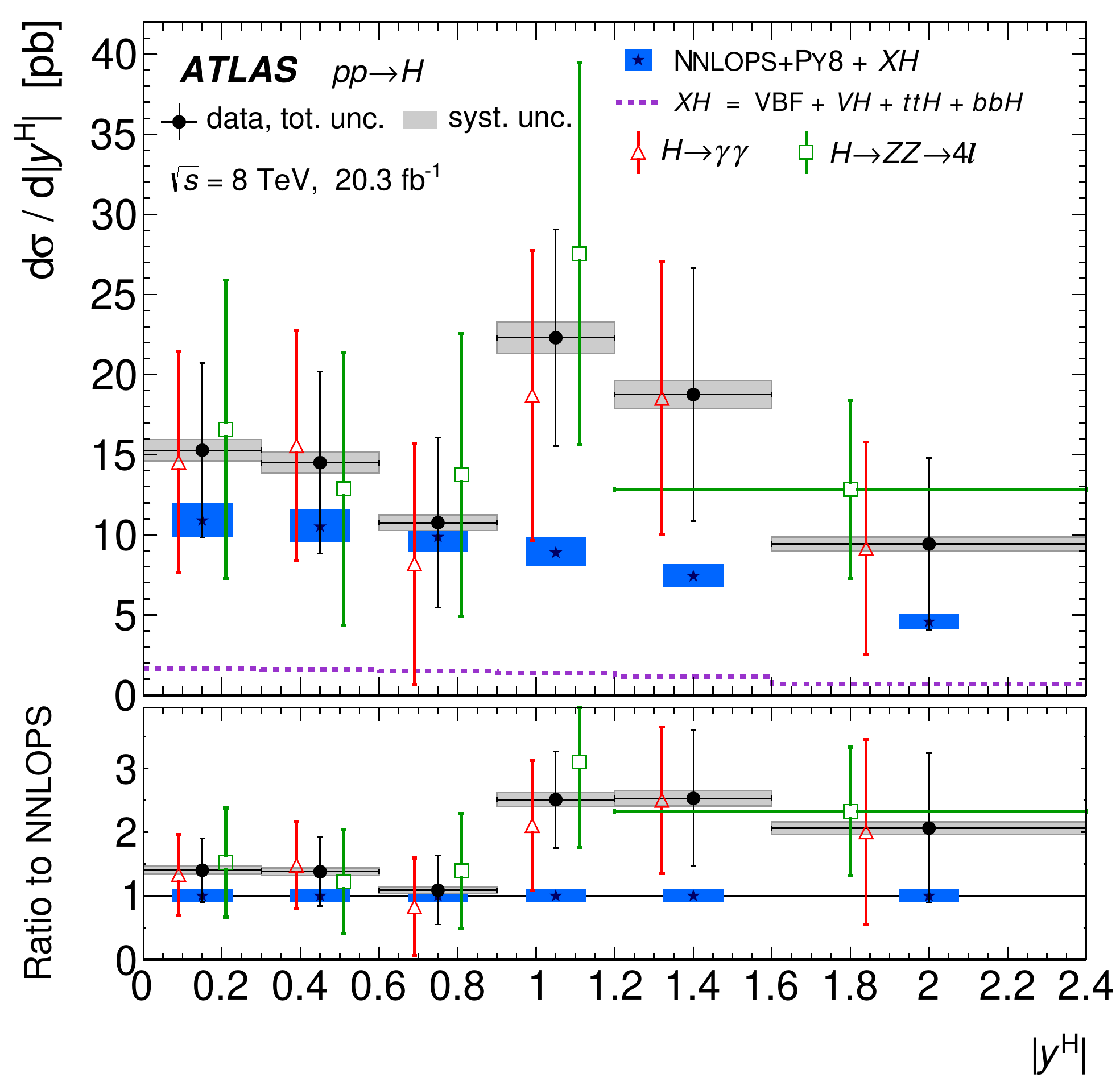}
\includegraphics[width=0.42\linewidth]{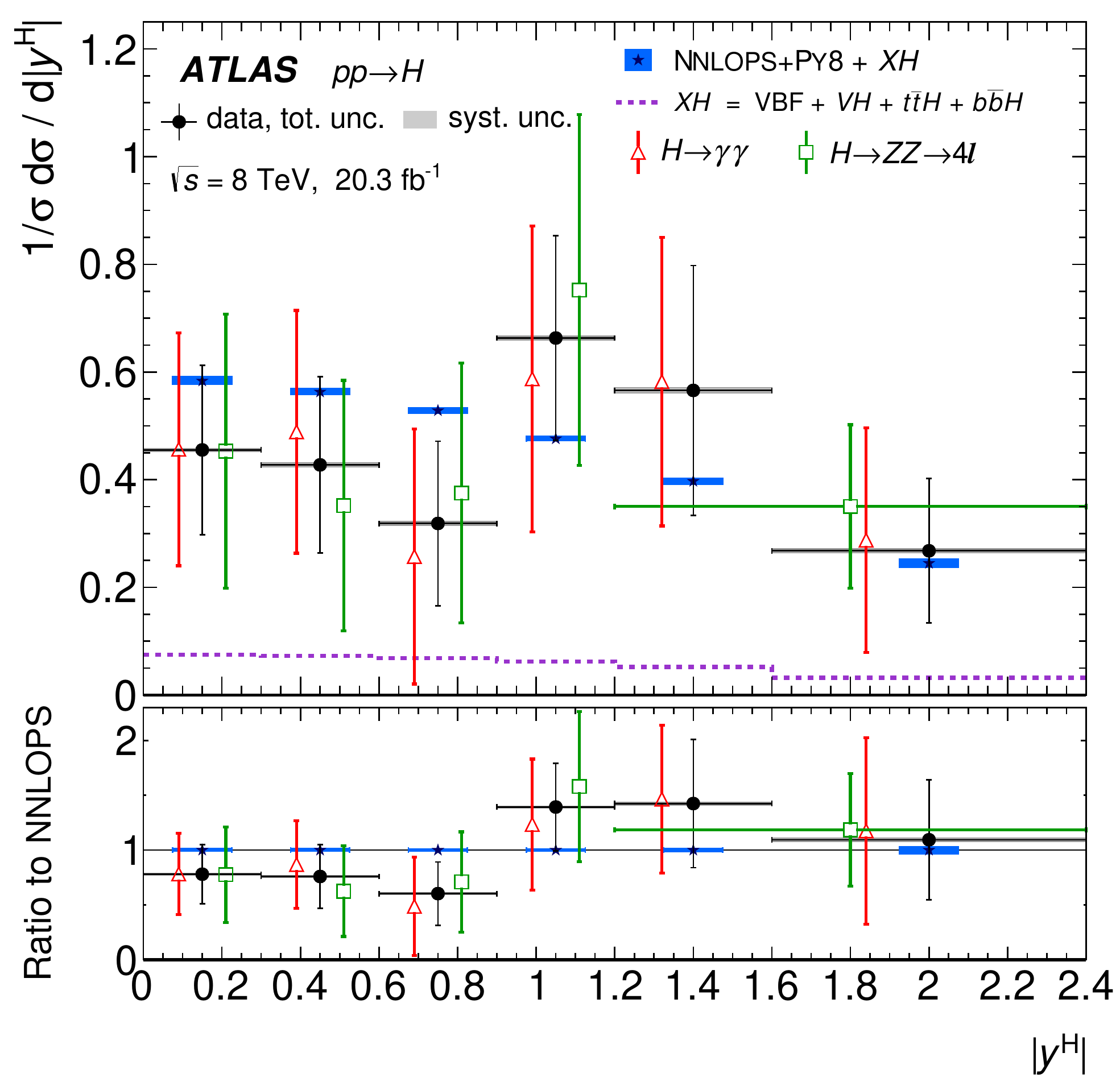}
\includegraphics[width=0.42\linewidth]{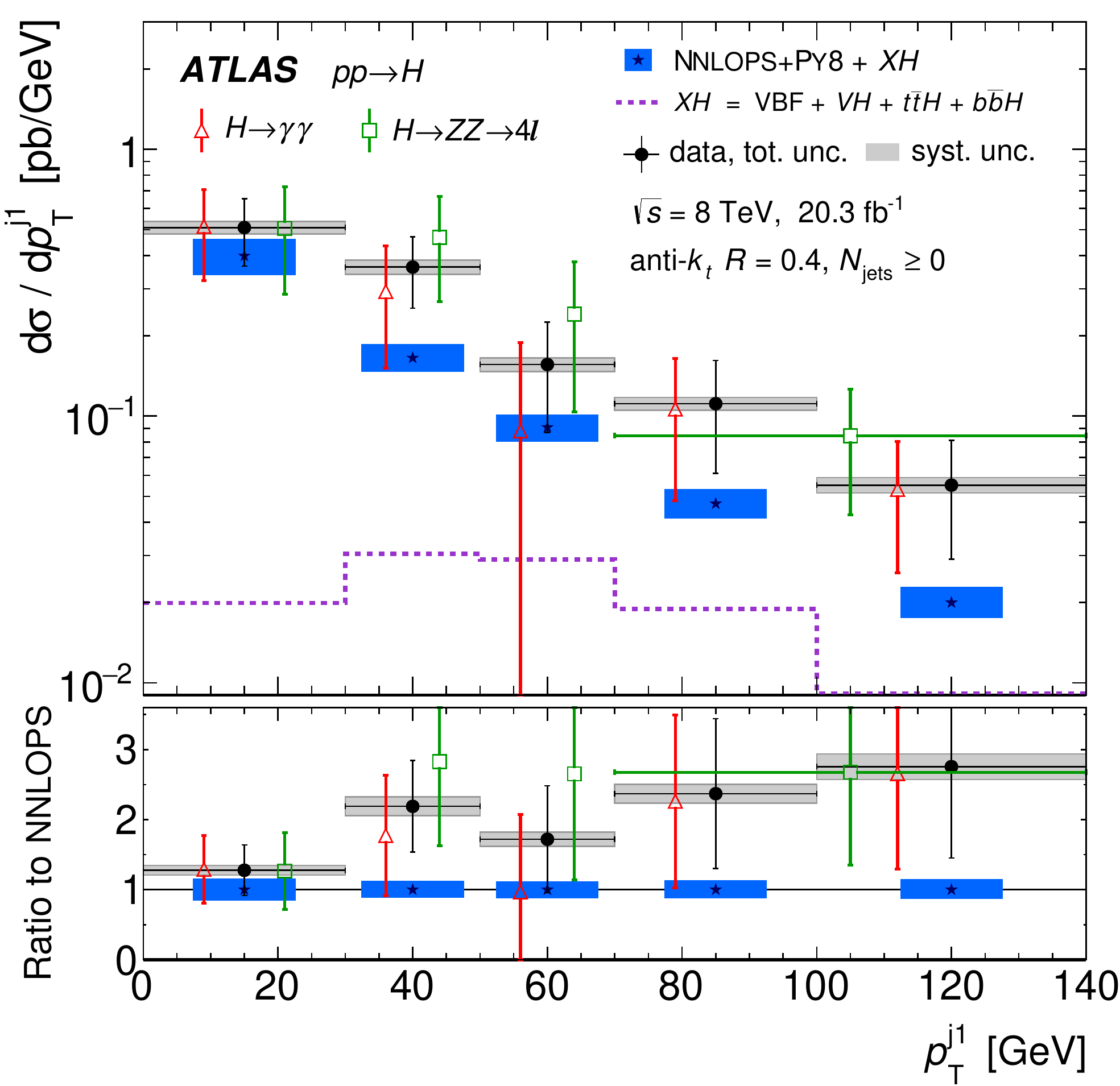}
\includegraphics[width=0.42\linewidth]{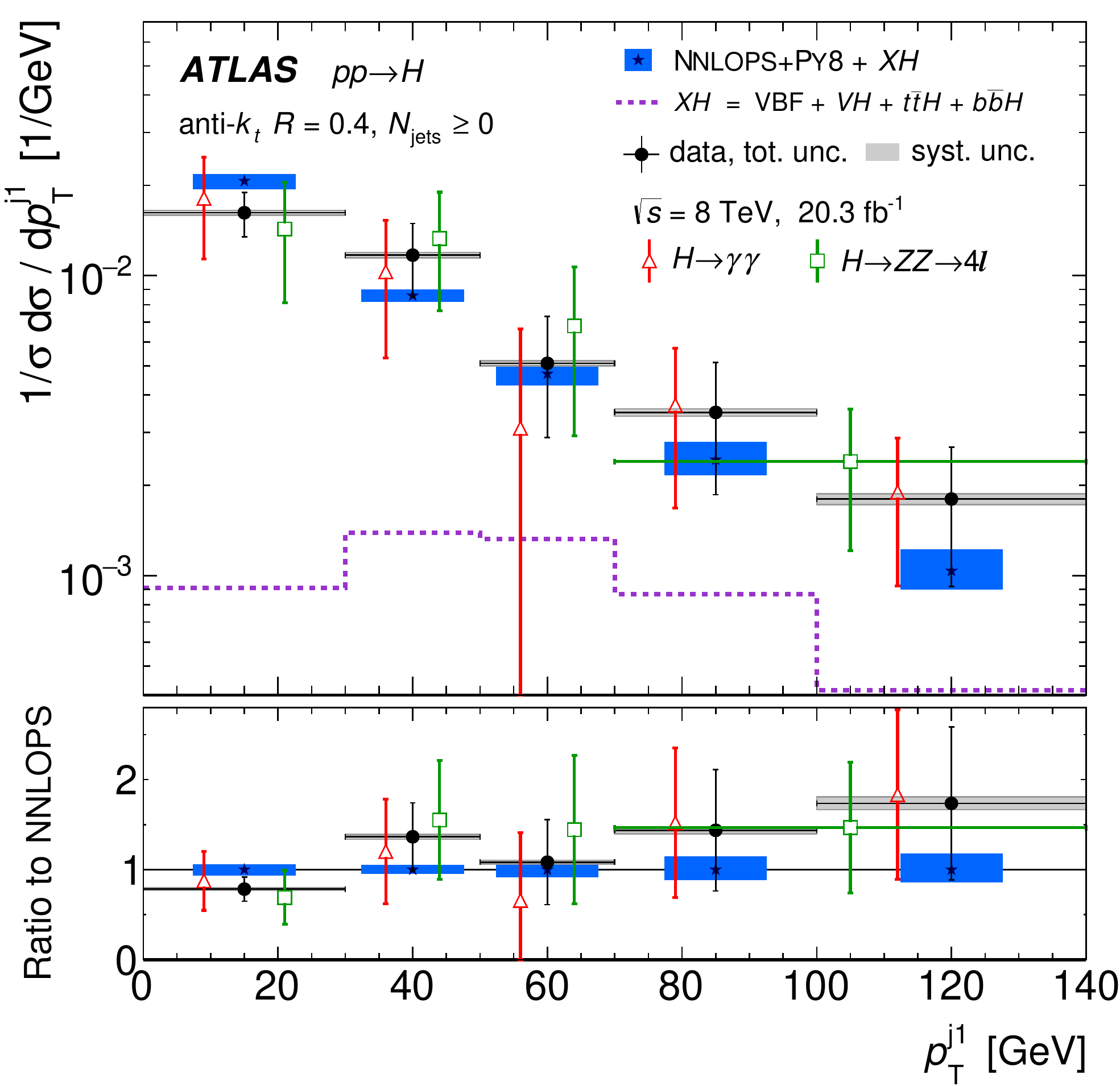}
\caption{\label{fig:xsecWchannels}
Differential cross sections (left) and shapes (right) of the Higgs boson transverse momentum (top), absolute rapidity (middle) and leading jet transverse momentum (bottom) of inclusive Higgs boson production at $\sqrt{s}=8$~TeV measured in the $\hyy{}$ and $\hzz{}$ final states using $20.3$~\ifb{} of $pp$ collisions.
Both the combined measurements as well as the individual channels are shown.}
\end{figure*}

\clearpage

\subsection{Result tables}
Tables~\ref{supp:XS:ptH}--\ref{supp:XS:ptjet} present the measured differential cross sections and
Tables~\ref{supp:shape:ptH}--\ref{supp:shape:ptjet} report the corresponding shape measurements.

\bgroup
\def\arraystretch{1.3}
\begin{table}[h!]
\centering
\caption{Measured cross section in bins of \ptH. The first uncertainty is statistical, the second is systematic. \label{supp:XS:ptH}}
\begin{tabular}{rr@{\;$\pm$\;}r@{\;$\pm$\;}r} \hline \hline
Bins [GeV] & \multicolumn{3}{c}{d$\sigma$/d\ptH\ [pb/GeV]}\\
\hline
0--20 &  0.20&0.15&0.01\\
20--30 &  0.88&0.27&0.04\\
30--40 &  0.46&0.23&0.02\\
40--50 &  0.56&0.20&0.03\\
50--60 &  0.28&0.18&0.01\\
60--80 &  0.188&0.090&0.009\\
80--100 &  0.136&0.067&0.006\\
100--200 &   0.0214&0.0091&0.0010\\
\hline \hline
\end{tabular}
\end{table}
\egroup

\bgroup
\def\arraystretch{1.3}
\begin{table}[h!]
\centering
\caption{Measured cross section in bins of \yH. The first uncertainty is statistical, the second is systematic. \label{supp:XS:yH} }
\begin{tabular}{rr@{\;$\pm$\;}r@{\;$\pm$\;}r} \hline \hline
Bins~~~ & \multicolumn{3}{c}{d$\sigma$/d\yH\ [pb]} \\
\hline
0.0--0.3~~~ &  15.3&5.4&0.7\\
0.3--0.6~~~ &  14.5&5.6&0.6\\
0.6--0.9~~~ &  10.8&5.3&0.5\\
0.9--1.2~~~ &  22.3&6.7&1.0\\
1.2--1.6~~~ &  18.8&7.9&0.9\\
1.6--2.4~~~ &  9.4&5.3&0.4\\
\hline \hline
\end{tabular}
\end{table}
\egroup 

\bgroup
\def\arraystretch{1.3}
\begin{table}[h!]
\centering
\caption{Measured cross section in bins of \njets. The first uncertainty is statistical, the second is systematic. \label{supp:XS:njets}}
\begin{tabular}{rr@{\;$\pm$\;}r@{\;$\pm$\;}r} \hline \hline
Bins & \multicolumn{3}{c}{d$\sigma$/d\njets\ [pb]} \\
\hline
0 &  15.3&4.3&0.8\\
1 &  11.9&2.7&0.6\\
2 &  4.1&1.5&0.3\\
$\ge 3$ &  2.57&0.97&0.32\\
\hline \hline
\end{tabular}
\end{table}
\egroup 

\bgroup
\def\arraystretch{1.3}
\begin{table}[h!]
\centering
\caption{Measured cross section in bins of \ptjet. The first uncertainty is statistical, the second is systematic. \label{supp:XS:ptjet}.}
\begin{tabular}{rr@{\;$\pm$\;}r@{\;$\pm$\;}r} \hline \hline
Bins [GeV] & \multicolumn{3}{c}{d$\sigma$/d\ptjet\ [pb/GeV]} \\
\hline
0--30 &  0.51&0.14&0.03\\
30--50 &   0.36&0.11&0.02\\
50--70 &  0.156&0.069&0.009\\
70--100 &  0.111&0.050&0.006\\
100--140 &  0.055&0.026&0.004\\
\hline \hline
\end{tabular}
\end{table}
\egroup 

\bgroup
\def\arraystretch{1.3}
\begin{table}[h!]
\centering
\caption{Measured fractions in bins of \ptH. The first uncertainty is statistical, the second is systematic. \label{supp:shape:ptH}}
\begin{tabular}{rr@{\;$\pm$\;}r@{\;$\pm$\;}r} \hline \hline
Bins [GeV] &  \multicolumn{3}{c}{1/$\sigma$ d$\sigma$/d\ptH\ [1/GeV]}\\
\hline
0--20 &  0.0055&0.0036&0.0006\\
20--30 &  0.0258&0.0077&0.0003\\
30--40 &  0.0133&0.0068&0.0002\\
40--50 &  0.0166&0.0060&0.0002\\
50--60 &  0.0079&0.0053&0.0001\\
60--80 &  0.0055&0.0027&0.0001\\
80--100 &  0.0041&0.0021&0.0000\\
100--200 &  0.00060&0.00026&0.00001\\
\hline \hline
\end{tabular}
\end{table}
\egroup
 
\bgroup
\def\arraystretch{1.3}
\begin{table}[h!]
\centering
\caption{Measured fractions in bins of \yH. The first uncertainty is statistical, the second is systematic. \label{supp:shape:yH}}
\begin{tabular}{rr@{\;$\pm$\;}r@{\;$\pm$\;}r} \hline \hline
Bins~~~ & \multicolumn{3}{c}{1/$\sigma$ d$\sigma$/d\yH} \\
\hline
0.0--0.3~~~ &  0.46&0.16&0.00\\
0.3--0.6~~~ &  0.43&0.16&0.00\\
0.6--0.9~~~ &  0.32&0.15&0.00\\
0.9--1.2~~~ &  0.66&0.19&0.00\\
1.2--1.6~~~ &  0.57&0.23&0.00\\
1.6--2.4~~~ &  0.27&0.13&0.00\\
\hline \hline
\end{tabular}
\end{table}
\egroup

\bgroup
\def\arraystretch{1.3}
\begin{table}[h!]
\centering
\caption{Measured fractions in bins of \njets. The first uncertainty is statistical, the second is systematic. \label{supp:shape:njets}}
\begin{tabular}{rr@{\;$\pm$\;}r@{\;$\pm$\;}r} \hline \hline
Bins & \multicolumn{3}{c}{1/$\sigma$ d$\sigma$/d\njets} \\
\hline
0 &  0.447&0.078&0.010\\
1 &  0.353&0.071&0.005\\
2 &  0.123&0.043&0.003\\
$\ge 3$ &  0.077&0.029&0.007\\
\hline \hline
\end{tabular}
\end{table}
\egroup 

\bgroup
\def\arraystretch{1.3}
\begin{table}[h!]
\centering
\caption{Measured fractions in bins of \ptjet. The first uncertainty is statistical, the second is systematic. \label{supp:shape:ptjet}}
\begin{tabular}{rr@{\;$\pm$\;}r@{\;$\pm$\;}r} \hline \hline
Bins [GeV] &  \multicolumn{3}{c}{1/$\sigma$ d$\sigma$/d\ptjet\ [1/GeV]}\\
\hline
0--30 &  0.0162&0.0027&0.0003\\
30--50 &  0.0117&0.0032&0.0002\\
50--70 &  0.0051&0.0022&0.0001\\
70--100 &  0.0035&0.0016&0.0001\\
100--140 &  0.00180&0.00088&0.00007\\
\hline \hline
\end{tabular}
\end{table}
\egroup 

\clearpage

\subsection{Uncertainty correlation tables}
Tables~\ref{supp:XScov:ptH}--\ref{supp:XScov:ptjet} contain the correlation matrices of the differential cross section measurements and
Tables~\ref{supp:shapecov:ptH}--\ref{supp:shapecov:ptjet} those of the differential shape measurements.

\bgroup
\def\arraystretch{1.2}
\begin{table}[h!]
\centering
\caption{Correlation matrix for the total uncertainty of the differential cross-section measurement in bins of \ptH.\label{supp:XScov:ptH}}
\begin{tabular}{ l rrrr rrrr }
\hline \hline
 & Bin 1 & Bin 2 & Bin 3 & Bin 4 & Bin 5 & Bin 6 & Bin 7 & Bin 8 \\
Bin 1  &  1.00  &  0.01  &  0.01  &  0.01  &  0.00  &  0.01  &  0.01  &  0.01 \\
Bin 2  &  0.01  &  1.00  &  $-$0.13  &  $-$0.10  &  0.01  &  0.01  &  0.01  &  0.02 \\
Bin 3  &  0.01  &  $-$0.13  &  1.00  &  $-$0.08  &  0.01  &  0.01  &  0.01  &  0.01 \\
Bin 4  &  0.01  &  $-$0.10  &  $-$0.08  &  1.00  &  0.01  &  0.01  &  0.01  &  0.01 \\
Bin 5  &  0.00  &  0.01  &  0.01  &  0.01  &  1.00  &  $-$0.16  &  $-$0.11  &  0.01 \\
Bin 6  &  0.01  &  0.01  &  0.01  &  0.01  &  $-$0.16  &  1.00  &  $-$0.11  &  0.01 \\
Bin 7  &  0.01  &  0.01  &  0.01  &  0.01  &  $-$0.11  &  $-$0.11  &  1.00  &  0.01 \\
Bin 8  &  0.01  &  0.02  &  0.01  &  0.01  &  0.01  &  0.01  &  0.01  &  1.00 \\
\hline \hline
\end{tabular}
\end{table}
\egroup

\bgroup
\def\arraystretch{1.2}
\begin{table}[h!]
\centering
\caption{Correlation matrix for the total uncertainty of the differential cross-section measurement in bins of \yH.\label{supp:XScov:yH}}
\begin{tabular}{ l rrrrrr }
\hline \hline
 & Bin 1 & Bin 2 & Bin 3 & Bin 4 & Bin 5 & Bin 6 \\
Bin 1  &  1.00  &  0.01  &  0.01  &  0.02  &  0.01  &  0.01 \\
Bin 2  &  0.01  &  1.00  &  0.01  &  0.02  &  0.01  &  0.01 \\
Bin 3  &  0.01  &  0.01  &  1.00  &  0.01  &  0.01  &  0.01 \\
Bin 4  &  0.02  &  0.02  &  0.01  &  1.00  &  0.02  &  0.01 \\
Bin 5  &  0.01  &  0.01  &  0.01  &  0.02  &  1.00  &  $-$0.28 \\
Bin 6  &  0.01  &  0.01  &  0.01  &  0.01  &  $-$0.28  &  1.00 \\
\hline \hline
\end{tabular}
\end{table}
\egroup
 
\bgroup
\def\arraystretch{1.2}
\begin{table}[h!]
\centering
\caption{Correlation matrix for the total uncertainty of the differential cross-section measurement in bins of \njets.\label{supp:XScov:njets}}
\begin{tabular}{ l rrrr }
\hline \hline
 & Bin 1 & Bin 2 & Bin 3 & Bin 4 \\
Bin 1  &  1.00  &  0.03  &  $-$0.02  &  0.02 \\
Bin 2  &  0.03  &  1.00  &  $-$0.04  &  0.05 \\
Bin 3  &  $-$0.02  &  $-$0.04  &  1.00  &  $-$0.04 \\
Bin 4  &  0.02  &  0.05  &  $-$0.04  &  1.00 \\
\hline \hline
\end{tabular}
\end{table}
\egroup 
 
\bgroup
\def\arraystretch{1.2}
\begin{table}[h!]
\centering
\caption{Correlation matrix for the total uncertainty of the differential cross-section measurement in bins of \ptjet.\label{supp:XScov:ptjet}}
\begin{tabular}{ l rrrrr }
\hline \hline
 & Bin 1 & Bin 2 & Bin 3 & Bin 4 & Bin 5 \\
Bin 1  &  1.00  &  0.03  &  0.02  &  0.02  &  0.01 \\
Bin 2  &  0.03  &  1.00  &  0.02  &  0.02  &  0.02 \\
Bin 3  &  0.02  &  0.02  &  1.00  &  0.02  &  0.01 \\
Bin 4  &  0.02  &  0.02  &  0.02  &  1.00  &  $-$0.19 \\
Bin 5  &  0.01  &  0.02  &  0.01  &  $-$0.19  &  1.00 \\
\hline \hline
\end{tabular}
\end{table}
\egroup

\bgroup
\def\arraystretch{1.2}
\begin{table}[h!]
\centering
\caption{Correlation matrix for the total uncertainty of the differential shape measurement in bins of \ptH.\label{supp:shapecov:ptH}}
\begin{tabular}{ l rrrrrrrr }
\hline \hline
 & Bin 1 & Bin 2 & Bin 3 & Bin 4 & Bin 5 & Bin 6 & Bin 7 & Bin 8 \\
Bin 1  &  1.00  &  $-$0.28  &  $-$0.22  &  $-$0.18  &  $-$0.16  &  $-$0.17  &  $-$0.11  &  $-$0.12 \\
Bin 2  &  $-$0.28  &  1.00  &  $-$0.34  &  $-$0.28  &  $-$0.11  &  $-$0.11  &  $-$0.08  &  $-$0.08 \\
Bin 3  &  $-$0.22  &  $-$0.34  &  1.00  &  $-$0.21  &  $-$0.08  &  $-$0.09  &  $-$0.06  &  $-$0.06 \\
Bin 4  &  $-$0.18  &  $-$0.28  &  $-$0.21  &  1.00  &  $-$0.07  &  $-$0.07  &  $-$0.05  &  $-$0.05 \\
Bin 5  &  $-$0.16  &  $-$0.11  &  $-$0.08  &  $-$0.07  &  1.00  &  $-$0.26  &  $-$0.18  &  $-$0.04 \\
Bin 6  &  $-$0.17  &  $-$0.11  &  $-$0.09  &  $-$0.07  &  $-$0.26  &  1.00  &  $-$0.18  &  $-$0.05 \\
Bin 7  &  $-$0.11  &  $-$0.08  &  $-$0.06  &  $-$0.05  &  $-$0.18  &  $-$0.18  &  1.00  &  $-$0.03 \\
Bin 8  &  $-$0.12  &  $-$0.08  &  $-$0.06  &  $-$0.05  &  $-$0.04  &  $-$0.05  &  $-$0.03  &  1.00 \\
\hline \hline
\end{tabular}
\end{table}
\egroup

\bgroup
\def\arraystretch{1.2}
\begin{table}[h!]
\centering
\caption{Correlation matrix for the total uncertainty of the differential shape measurement in bins of \yH.\label{supp:shapecov:yH}}
\begin{tabular}{ l rrrrrr }
\hline \hline
 & Bin 1 & Bin 2 & Bin 3 & Bin 4 & Bin 5 & Bin 6 \\
Bin 1  &  1.00  &  $-$0.09  &  $-$0.09  &  $-$0.11  &  $-$0.10  &  $-$0.21 \\
Bin 2  &  $-$0.09  &  1.00  &  $-$0.09  &  $-$0.12  &  $-$0.11  &  $-$0.22 \\
Bin 3  &  $-$0.09  &  $-$0.09  &  1.00  &  $-$0.11  &  $-$0.10  &  $-$0.21 \\
Bin 4  &  $-$0.11  &  $-$0.12  &  $-$0.11  &  1.00  &  $-$0.13  &  $-$0.27 \\
Bin 5  &  $-$0.10  &  $-$0.11  &  $-$0.10  &  $-$0.13  &  1.00  &  $-$0.66 \\
Bin 6  &  $-$0.21  &  $-$0.22  &  $-$0.21  &  $-$0.27  &  $-$0.66  &  1.00 \\
\hline \hline
\end{tabular}
\end{table}
\egroup 
 
\bgroup
\def\arraystretch{1.2}
\begin{table}[h!]
\centering
\caption{Correlation matrix for the total uncertainty of the differential shape measurement in bins of \njets.\label{supp:shapecov:njets}}
\begin{tabular}{ l rrrr }
\hline \hline
 & Bin 1 & Bin 2 & Bin 3 & Bin 4 \\
Bin 1  &  1.00  &  $-$0.77  &  $-$0.37  &  $-$0.26 \\
Bin 2  &  $-$0.77  &  1.00  &  $-$0.16  &  $-$0.11 \\
Bin 3  &  $-$0.37  &  $-$0.16  &  1.00  &  $-$0.05 \\
Bin 4  &  $-$0.26  &  $-$0.11  &  $-$0.05  &  1.00 \\
\hline \hline
\end{tabular}
\end{table}
\egroup 

\bgroup
\def\arraystretch{1.2}
\begin{table}[h!]
\centering
\caption{Correlation matrix for the total uncertainty of the differential shape measurement in bins of \ptjet.\label{supp:shapecov:ptjet}}
\begin{tabular}{ lrrrrr }
\hline \hline
 & Bin 1 & Bin 2 & Bin 3 & Bin 4 & Bin 5 \\
Bin 1  &  1.00  &  $-$0.63  &  $-$0.38  &  $-$0.36  &  $-$0.20 \\
Bin 2  &  $-$0.63  &  1.00  &  $-$0.12  &  $-$0.11  &  $-$0.06 \\
Bin 3  &  $-$0.38  &  $-$0.12  &  1.00  &  $-$0.07  &  $-$0.04 \\
Bin 4  &  $-$0.36  &  $-$0.11  &  $-$0.07  &  1.00  &  $-$0.28 \\
Bin 5  &  $-$0.20  &  $-$0.06  &  $-$0.04  &  $-$0.28  &  1.00 \\
\hline \hline
\end{tabular}
\end{table}
\egroup

\clearpage

\subsection{Gluon fusion cross section}

Figure~\ref{fig:totHiggsXsec} shows the measurement of the  Higgs boson production cross section compared to 
a range of theory predictions, including LHC-XS, the result used by the ATLAS and CMS collaboration in Run 1, 
for which the ggF part is accurate to NNLO+NNLL in QCD~\cite{LHCHiggsCrossSectionWorkingGroup:2011ti},
as well as ggF cross section calculations that attempt to go beyond NNLO, including the recently completed full N$^3$LO
prediction. Details about the various predictions are presented in Table~\ref{tab:TheoryIncl},
and the central values and a breakdown of the uncertainties of the calculations
as well as the measurement are reported in Table~\ref{tab:ggFinclValues}.

\begin{figure}[h!]
\centering
\includegraphics[width=0.98\linewidth]{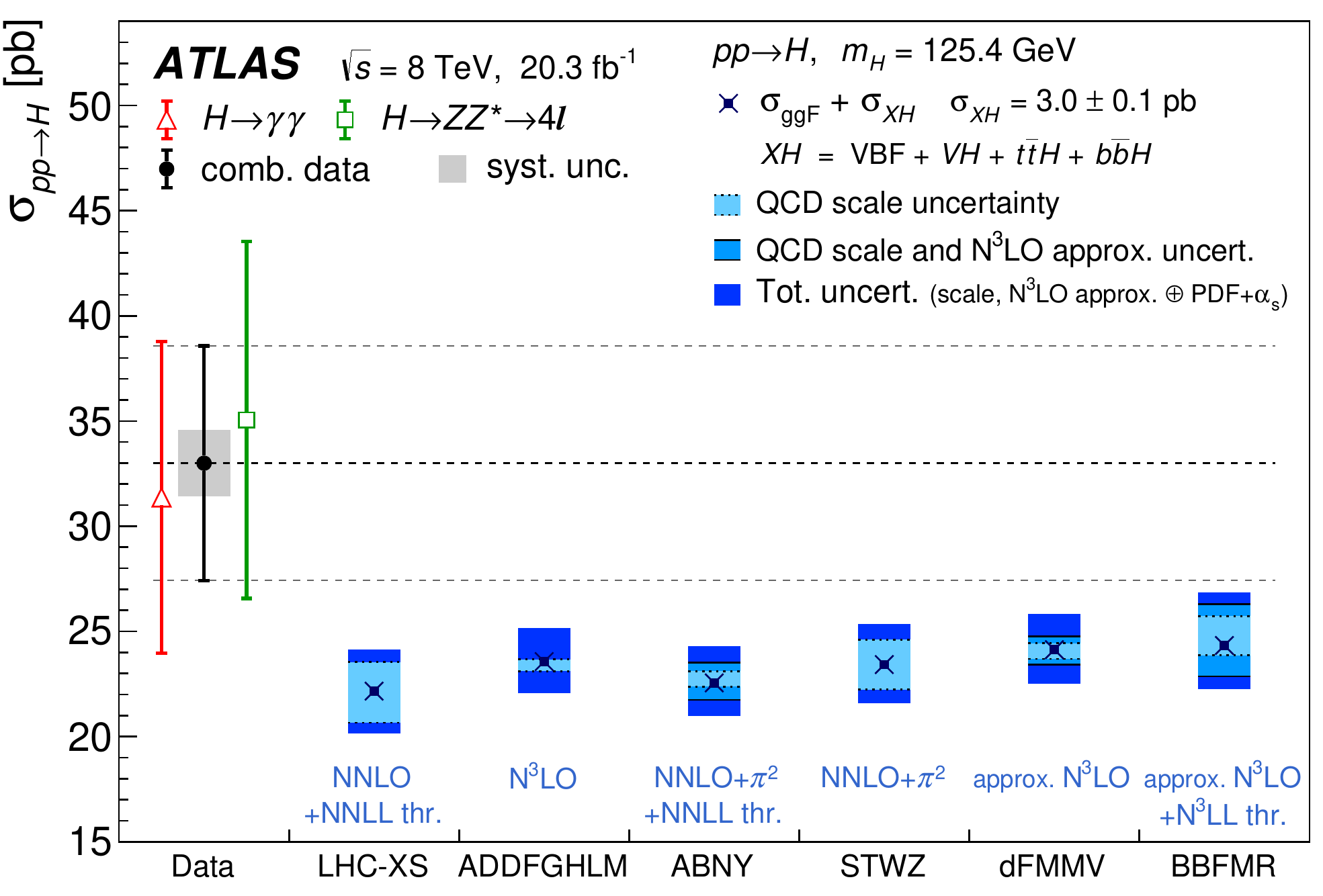}
\caption{\label{fig:totHiggsXsec}
Measured total cross section of Higgs boson production compared to different theoretical calculations.
}
\end{figure}

\bgroup
\def\arraystretch{1.2}
\begin{table}[h!]
\centering
\caption{Summary of the ggF predictions used in the comparison with the measured cross sections.
The second column states the order in QCD perturbation theory and which threshold resummation is applied, if any.
Further details are provided in the footnotes.
All predictions are for $m_H=125.4$~GeV and $\sqrt{s}=8$~TeV.
\label{tab:TheoryIncl}}
\begin{tabular}{ll}
\hline \hline
\multicolumn{2}{c}{Total cross-section calculations} \tabularnewline
\hline
  LHC-XS~\cite{LHCHiggsCrossSectionWorkingGroup:2011ti} & NNLO+NNLL$^{\mathrm{a,b,c}}$\\
  ADDFGHLM~\cite{Anastasiou:2011pi,Anastasiou:2014vaa,Anastasiou:2014lda,Anastasiou:2015ema}  & N$^{3}$LO$^{\mathrm{a,b,c}}$\\
ABNY~\cite{Ahrens:2010rs}           & NNLO+NNLL$^{\mathrm{a,b,c,d,e}}$\\
STWZ~\cite{Stewart:2013faa}         & NNLO$^{\mathrm{c,d}}$\\
dFMMV~\cite{deFlorian:2014vta}      & approx. N$^{3}$LO$^{\mathrm{c}}$\\
BBFMR~\cite{Ball:2013bra,Bonvini:2014jma,Bonvini:2014joa} & approx. N$^{3}$LO+N$^{3}$LL$^{\mathrm{a,b,c}}$\\
\hline
\hline
\end{tabular}
\begin{footnotesize}
\def\arraystretch{1.0}
\begin{tabular}{ll}
$^{\mathrm{a}}$ & Considers $b$- (and $c$-) quark masses in the $gg\to H$ loop\\
$^{\mathrm{b}}$ & Includes electroweak corrections\\
$^{\mathrm{c}}$ & Based on MSTW2008nnlo~\cite{Martin:2009iq} ($\alpha_{\mathrm{s}}$ from PDF set)\\
$^{\mathrm{d}}$ & Uses $\pi^2$-resummed $gg\to H$ form factor\\
$^{\mathrm{e}}$ & In the counting of Ref.~\cite{Ahrens:2010rs}, the result has N$^3$LL accuracy
\end{tabular}
\end{footnotesize}
\end{table}
\egroup

\bgroup
\def\arraystretch{1.5}
\begin{table}[h!]
\centering
\caption{Central values and uncertainties for the different ggF predictions and the data. \label{tab:ggFinclValues}}
\begin{tabular}{lrll} \hline \hline
Name &  \multicolumn{3}{c}{$\sigma_{gg\to H}$ [pb]}\\
\hline
Data$-XH$\footnote{Non-ggF cross section $\sigma_{XH} = 3.01 ^{+0.05}_{-0.06}\ \mathrm{(scale)}\ \pm 0.09\ \mathrm{(pdf)}$~pb, subtracted from the measured inclusive cross section: $33.0 \pm 5.3 \ \mathrm{(stat)} \pm 1.6 \ \mathrm{(sys)}$~pb.} & 30.0 & $\pm 5.3 \ \mathrm{(stat)}$ &  $\pm 1.6\ \mathrm{(sys)}$\\
\hline
LHC-XS             & 19.15 & $^{+1.38}_{-1.49}$\ (scale) & $^{+1.44}_{-1.32}$\ (pdf)\\
ADDFGHLM           & 20.55 & $^{+0.04}_{-0.45}$\ (scale) & $^{+1.60}_{-1.44}$\ (pdf)\\
ABNY               & 19.54 & $^{+0.55}_{-0.14}$\ (scale) & $^{+1.47}_{-1.35}$\ (pdf)\ $\pm 0.78$\ (appr.)\\
STWZ               & 20.41 & $\pm 1.18$\ (scale)         & $^{+1.53}_{-1.41}$\ (pdf)\\
dFMMV              & 21.12 & $^{+0.29}_{-0.42}$\ (scale) & $^{+1.58}_{-1.46}$\ (pdf)\ $\pm 0.56$\ (appr.)\\
BBFMR              & 21.32 & $^{+1.39}_{-0.45}$\ (scale) & $^{+1.60}_{-1.47}$\ (pdf) $\pm 1.39$\ (appr.)\\
\hline \hline
\end{tabular}
\end{table}
\egroup

For the predictions, uncertainties from renormalization, factorization and, where appropriate, resummation scale variations as well as uncertainties due to approximation or missing terms beyond NNLO are provided separately for each prediction.
The same relative PDF uncertainty of $^{+7.5}_{-6.9}$\% is assigned to all ggF predictions, except for the ADDFGHLM prediction for which this uncertainty is increased to $^{+7.8}_{-7.0}$\% corresponding to the change in MSTW2008nnlo uncertainty observed by the group when changing the matrix element from NNLO to N$^{3}$LO.
The non-ggF contribution ($\sigma_{XH} = 3.01 ^{+0.05}_{-0.06}\ \mathrm{(scale)}\ \pm 0.09\ \mathrm{(pdf)}$~pb, $XH$ = VBF + $VH$ + $t\bar{t}H$ + $b\bar{b}H$)
is added to the ggF predictions to be able to compare to the data in Fig.~\ref{fig:totHiggsXsec}.

As detailed in Table~\ref{tab:TheoryIncl}, all inclusive predictions use the same PDF set but differ in the perturbative calculation.
Four of the predictions apply both electroweak corrections and consider finite $b$- and $c$-quark masses.
These corrections have non-negligible impacts on the ggF cross section;
the electroweak correction results in an increase of approximately 5\%, while
the bottom and charm corrections give a $O(5-10\%)$ reduction depending on precisely how they have been implemented in the calculations.
They therefore have an opposite effect on the total cross section such that their numerical effects partially cancel.
The STWZ and dFMMV predictions consider neither of these corrections.

The calculations take different approaches to approximately evaluate the ggF cross section beyond NNLO.
Therefore the preferred scale for each calculation differs,
and the choice of scale and the precise scale variations applied was left to the authors of the calculations.
The LHC-XS, ABNY, STWZ, and BBFMR predictions use a central scale of $\mu_0=m_H$ as their overall scale,
while dFMMV and ADDFGHLM use $\mu_0=m_H/2$.

\clearpage
\subsection{Compatibility between predictions and data}

Tables~\ref{tab:pvaluesdiff} and~\ref{tab:CompatibilityTheory} present compatibility tests between the differential predictions and the measured cross sections and shapes respectively. The theory uncertainties are assumed to be Gaussian and to be fully correlated between bins.

\bgroup
\def\arraystretch{1.2}
\begin{table}[htb]
\centering
\caption{$p$-values quantifying the compatibility between predictions and the data for
the differential cross sections. The theory uncertainties are assumed to be Gaussian and to be fully correlated between bins. \label{tab:pvaluesdiff}}
\begin{tabular}{lccc}
\hline
\hline
 & \ptH & \yH &  \ptjet \\
\hline
\HRes & 2\% &14\%  & - \\
STWZ &- & - &  26\%  \\
JetVHeto & - & - & 24\%\\ \hline
\hline
\end{tabular}
\end{table}
\egroup

\bgroup
\def\arraystretch{1.2}
\begin{table}[htb]
\centering
\caption{$p$-values quantifying the compatibility between predictions and data for the differential shapes.
The theory uncertainties are assumed to be Gaussian distributed and fully correlated between bins. \label{tab:CompatibilityTheory}}
\begin{tabular}{lccc}
\hline
\hline
 & \ptH & \yH &  \ptjet \\
\hline
\HRes & 15\% & 64\% &  - \\
NNLOPS & 10\% & 64\% &  64\% \\
SHERPA 2.1.1 & 22\% & 63\% &  88\% \\
MG5\_aMC@NLO & 8\% & 60\% & 88\% \\
\hline
\hline
\end{tabular}
\end{table}
\egroup

\subsection{Non-perturbative correction factors}

Table~\ref{tab:NPcorr} presents multiplicative non-perturbative correction factors and associated uncertainties that are applied to correct analytical parton-level predictions presented in this Letter to particle level.
These corrections account for hadronization and multiple parton interactions, and are derived based on a number of underlying event and showering tunes applied to the Higgs boson production MC samples used in the analysis.

\bgroup
\def\arraystretch{1.2}
\begin{table}[hbt]
\begin{center}
\caption{
Non-perturbative factors in percent with systematic uncertainties, accounting for the impact of hadronization and underlying event.
\label{tab:NPcorr} }
\vspace{0.2cm}
\begin{tabular}{lrrrrrrrr}
\hline\hline
Bin & \multicolumn{1}{c}{1} & \multicolumn{1}{c}{2} & \multicolumn{1}{c}{3} & \multicolumn{1}{c}{4} & \multicolumn{1}{c}{5} & \multicolumn{1}{c}{6} & \multicolumn{1}{c}{7} & \multicolumn{1}{c}{8} \\
\hline
\ptH          & $ 99.5\pm 1.0$ & $100.5\pm 1.3$ & $99.9\pm 1.1$ & $ 99.9\pm 1.4$ & $100.2\pm 1.4$ & $99.9\pm 1.1$ & $100.4\pm 1.5$ & $100.0\pm 1.0$ \\
\yH           & $100.0\pm 0.8$ & $100.1\pm 0.7$ & $99.8\pm 0.7$ & $100.1\pm 0.8$ & $100.0\pm 0.7$ & $99.8\pm 0.6$ & & \\
\njets, excl  & $100.6\pm 2.1$ & $ 99.8\pm 2.9$ & $99.3\pm 5.0$ & $ 96.7\pm 6.7$ & & & \\
\njets, incl  & $100.0\pm 0.0$ & $ 99.2\pm 2.5$ & $98.7\pm 5.3$ & $ 96.8\pm 6.7$ & & & \\
\ptjet        & $100.6\pm 2.1$ & $ 99.6\pm 2.6$ & $99.2\pm 2.9$ & $ 99.1\pm 3.2$ & $ 98.7\pm 3.6$ & & &\\
\hline\hline
\end{tabular}
\end{center}
\end{table}
\egroup

\clearpage


\begin{thebibliography}{10}
\bibitem{ATLASCoordinateSystem}
 ATLAS uses a right-handed coordinate system with its origin at the nominal
  interaction point (IP) in the center of the detector and the $z$-axis along
  the beam pipe. The $x$-axis points from the IP to the center of the LHC ring,
  and the $y$-axis points upward. Cylindrical coordinates $(r,\phi )$ are used
  in the transverse plane, $\phi $ being the azimuthal angle around the beam
  pipe. The pseudorapidity is defined in terms of the polar angle $\theta $ as
  $\eta = - 2 \ln \tan(\theta/2)$.
\end{thebibliography}
\end{document}